\documentclass[useAMS,usenatbib]{mnras}
\usepackage{graphicx}
\usepackage[hang]{subfigure}
\usepackage{ amssymb }
\usepackage{color}
\usepackage{ amsmath }
\usepackage{fmtcount}
\usepackage{array,multirow}
\usepackage{booktabs}
\usepackage{appendix}

\DeclareRobustCommand{\Cunha}[3]{#2}
\let\Cunhathebibliography\thebibliography
\def\thebibliography{\DeclareRobustCommand{\Cunha}[3]{##3}\Cunhathebibliography}

\DeclareRobustCommand{\Barros}[3]{#2}
\let\Barrosthebibliography\thebibliography
\def\thebibliography{\DeclareRobustCommand{\Barros}[3]{##3}\Barrosthebibliography}

\usepackage[dvipsnames]{xcolor}
\newcommand{\txn}[1]{\textnormal{#1}}

\newcommand{\ltsim}{\mbox{{\raisebox{-0.4ex}{$\stackrel{<}{{\scriptstyle\sim}}$}}}}

\newcommand{\HST}{\hbox{\emph{HST}}}
\newcommand{\JWST}{\hbox{\emph{JWST}}}
\newcommand{\Spitzer}{\hbox{\emph{Spitzer}}}
\newcommand{\beagle}{\textsc{beagle}}
\newcommand{\leopy}{\textsc{LEO-Py}}
\newcommand{\astrodeep}{\textsc{astrodeep}}

\newcommand{\sextractor}{\textsc{sextractor}}

\newcommand{\tphot}{\textsc{tphot}}
\newcommand{\kelly}{K07}
\newcommand{\CL}{CL21}

\newcommand{\Mdist}{\hbox{$\boldsymbol{\eta}$}}

\newcommand{\zdist}{\hbox{$\boldsymbol{\theta}$}}

\newcommand{\yr}{\hbox{$\mathrm{yr}$}}
\newcommand{\Myr}{\hbox{$\mathrm{Myr}$}}
\newcommand{\cm}{\hbox{$\mathrm{cm}$}}
\newcommand{\A}{\hbox{$\txn{\AA}$}}
\newcommand{\um}{\hbox{$\mathrm{\mu m}$}}
\newcommand{\Msun}{\hbox{$\mathrm{M_{\odot}}$}}
\newcommand{\Zsun}{\hbox{$\mathrm{Z_{\odot}}$}}
\newcommand{\COsun}{\hbox{$\mathrm{(C/O)_{\odot}}$}}

\newcommand{\Mup}{\hbox{$M_\mathrm{up}$}}

\newcommand{\Z}{\hbox{$Z$}}
\newcommand{\Zism}{\hbox{$Z_\textsc{ism}$}}
\newcommand{\msa}{\hbox{$t$}}
\newcommand{\tausfr}{\hbox{$\tau_\textsc{sfr}$}}
\newcommand{\logmsa}{\hbox{$\log(\msa/\yr)$}}
\newcommand{\logtausfr}{\hbox{$\log(\tausfr/\yr)$}}
\newcommand{\tauV}{\hbox{$\hat{\tau}_\textsc{v}$}}
\newcommand{\tauVism}{\hbox{$\hat{\tau}^\textsc{ism}_\textsc{v}$}}
\newcommand{\tauVbc}{\hbox{$\hat{\tau}^\textsc{bc}_\textsc{v}$}}
\newcommand{\Us}{\hbox{$U_\textsc{s}$}}
\newcommand{\logUs}{\hbox{$\log\Us$}}
\newcommand{\xid}{\hbox{$\xi_\mathrm{d}$}}
\newcommand{\mud}{\hbox{$\mu_\mathrm{d}$}}
\newcommand{\nh}{\hbox{$n_\textsc{h}$}}
\newcommand{\CO}{\hbox{$\mathrm{(C/O)}$}}

\newcommand{\hii}{\hbox{$\mathrm{H\,\textsc{ii}}$}}

\newcommand{\Mstar}{\hbox{$\mathrm{M_{\star}}$}}
\newcommand{\MstarInLog}{\hbox{$M$}}
\newcommand{\MtotInLog}{\hbox{$M_\mathrm{tot}$}}
\newcommand{\sfr}{\hbox{${\Psi}$}} 
\newcommand{\sfrInLog}{\hbox{${\psi}$}}

\newcommand{\redshift}{\hbox{$z$}}

\newcommand{\MstarSFR}{\hbox{$\Mstar\mathrm{-}\sfr$}}

\newcommand{\MstarSFRredshift}{\hbox{$\Mstar\mathrm{-}\sfr\mathrm{-}\redshift$}}

\newcommand{\slope}{\hbox{$\beta$}}
\newcommand{\intercept}{\hbox{$\alpha_\mathrm{9.7}$}}
\newcommand{\scatter}{\hbox{$\sigma$}}

\newcommand{\ssfrNorm}{\hbox{$N$}}
\newcommand{\ssfrPower}{\hbox{$\gamma$}}

\newcommand{\OLmean}{\hbox{$\mu_\textsc{OL}$}}
\newcommand{\OLscatter}{\hbox{$\sigma_\textsc{OL}$}}
\newcommand{\OLprob}{\hbox{$p_\textsc{OL}$}}
\newcommand{\PMS}{\hbox{$\mathrm{P}_\textsc{MS}$}}
\newcommand{\POL}{\hbox{$\mathrm{P}_\textsc{OL}$}}

\newcommand{\prob}{\hbox{$\mathrm{P}$}}
\newcommand{\conditional}[2]{\hbox{$\prob(#1 \mid #2)$}}
\newcommand{\ThetaB}{\hbox{$\mathbf{\Theta}$}}
\newcommand{\DB}{\hbox{$\mathbf{D}$}}
\newcommand{\totsample}{$1038$}
\newcommand{\OLhogg}{\hbox{OL-Gauss}}
\newcommand{\OLnone}{\hbox{OL-Minimal}}
\newcommand{\OLclipped}{\hbox{OL-Clipped}}

\newcommand{\oii}{\hbox{$[\txn{O}\textsc{ii}]\lambda\lambda3726\txn{\AA},3729\txn{\AA}$}}
\newcommand{\oiii}{\hbox{$[\txn{O}\textsc{iii}]\lambda5007\txn{\AA}$}}
\newcommand{\Hb}{\hbox{$\mathrm{H}\beta$}}
\newcommand{\Ha}{\hbox{$\mathrm{H}\alpha$}}
\newcommand{\La}{\hbox{$\mathrm{Ly}\alpha$}}

\title[M$_{*}$-SFR]{Bayesian hierarchical modelling of the M$_{*}$-SFR relation from $1\ltsim z\ltsim6$ in ASTRODEEP}
\author[L. Sandles et al.]{L. Sandles$^{1,2}$\thanks{Email: ls861@cam.ac.uk}., E. Curtis-Lake$^{3}$, S. Charlot$^{4}$, J. Chevallard$^{5}$, R. Maiolino$^{1,2,6}$\\
$^{1}$ Kavli Institute for Cosmology, University of Cambridge, Madingley Road, Cambridge, CB3 0HA, UK\\
$^{2}$ Cavendish Laboratory Astrophysics Group, University of Cambridge, 19 JJ Thomson Avenue, Cambridge, CB3 0HE, UK\\
$^{3}$ Centre for Astrophysics Research, Department of Physics, Astronomy and Mathematics, University of Hertfordshire, \\Hatfield, AL10 9AB, UK\\
$^4$ Sorbonne Universit\'e, CNRS, UMR7095, Institut d'Astrophysique de Paris, F-75014, Paris, France\\
$^5$ Sub-department of Astrophysics, Department of Physics, University of Oxford, Denys Wilkinson Building, Keble Road, \\Oxford OX1 3RH, UK\\
$^6$Department of Physics and Astronomy, University College London, Gower Street, London WC1E 6BT, UK}

\begin{document}

\date{}

\pagerange{\pageref{firstpage}--\pageref{lastpage}} \pubyear{2002}

\maketitle

\label{firstpage}

\begin{abstract}
\noindent
The Hubble Frontier Fields represent the opportunity to probe the high-redshift evolution of the main sequence of star-forming galaxies to lower masses than possible in blank fields thanks to foreground lensing of massive galaxy clusters. We use the \beagle\ SED-fitting code to derive stellar masses, $\Mstar=\log(\MstarInLog/\Msun)$, SFRs, $\sfr=\log(\sfrInLog/\Msun\,\yr^{-1})$ and redshifts from galaxies within the \astrodeep\ catalogue. We fit a fully Bayesian hierarchical model of the main sequence over $1.25<z<6$ of the form $\sfr = \intercept(z) + \slope(\Mstar-9.7) + \mathcal{N}(0,\scatter^2)$ while explicitly modelling the outlier distribution.  The redshift-dependent intercept at $\Mstar=9.7$ is parametrized as $\intercept(z) = \log[\ssfrNorm (1+\redshift)^{\ssfrPower}] + 0.7$. Our results agree with an increase in normalization of the main sequence to high redshifts that follows the redshift-dependent rate of accretion of gas onto dark matter halos with $\ssfrPower=2.40^{+0.18}_{-0.18}$.  We measure a slope and intrinsic scatter of $\slope=0.79^{+0.03}_{-0.04}$ and $\scatter=0.26^{+0.02}_{-0.02}$. We find that the sampling of the SED provided by the combination of filters (\textit{Hubble} + ground-based Ks-band + \Spitzer{} 3.6 and 4.5 \um) is insufficient to constrain \Mstar\ and \sfr\ over the full dynamic range of the observed main sequence, even at the lowest redshifts studied. While this filter set represents the best current sampling of high-redshift galaxy SEDs out to $z>3$, measurements of the main sequence to low masses and high redshifts still strongly depend on priors employed in SED fitting (as well as other fitting assumptions).  Future data-sets with \JWST\ should improve this.
\end{abstract}


\newpage

\section{Introduction}

The relationship between star formation rate (SFR) and stellar mass of  ``normal'' star-forming galaxies has been well studied and is often referred to as the ``star-forming main sequence'' (originally labelled as such by \citealt{noeske_star_2007}). For masses less than $\log(\MstarInLog/\Msun)\lesssim10.1$, the main sequence is commonly parametrized as a straight line while at higher masses there is evidence for a redshift-dependent turn-over \citep{whitaker_constraining_2014,lee_turnover_2015,schreiber_herschel_2015,tasca_evolving_2015,tomczak_sfr-m_2016,leslie_vla-cosmos_2020,leja_new_2021}. ALMA observations suggest that the resolved main sequence is a by-product of two more physically connected relations; that between stellar mass and molecular gas densities, and that between the molecular gas and star formation rate densities \citep{lin_almaquest_2019, baker_almaquest_2022}.  However, direct measurements of the molecular gas reservoir are unfeasible for large samples at high redshifts, and measurements of the main sequence remain relevant as we move into the \textit{James Webb Space Telescope} (\JWST) era.

\cite{speagle_highly_2014} provide a thorough review of a compilation of 25 studies of the star-forming main sequence. They show that many of the discrepancies between measurements of slope and normalization can be resolved once two primary issues have been corrected for: the method chosen to select star-forming galaxies and the method used to calculate SFR (e.g. from emission lines, rest-frame ultra-violet continuum, spectral-energy distribution fitting). Having calibrated the results within the literature, \cite{speagle_highly_2014} report that both the slope ($\sim0.4 - 0.8$) and normalization ($\sim2$ orders of magnitude) increase from redshift 0 to 4, whilst the intrinsic scatter remains relatively constant ($\sim0.2$ dex). 

In recent years much work has been done to constrain the star-forming main sequence at higher redshifts \citep{steinhardt_star_2014, salmon_relation_2015, santini_star_2017, pearson_main_2018, thorne_deep_2021, bhatawdekar_uv_2021}. \cite{steinhardt_star_2014} show that for massive galaxies ($>10^{10}\,\Msun$) the MS extends to at least $z=6$. \cite{salmon_relation_2015} use multi-wavelength photometry to determine an almost constant main sequence relation, though with mildly increasing normalization, between $3.5<z<6.5$. They study samples chosen at constant number density spanning the redshift range to link progenitor galaxies, finding evidence for rising  star formation histories (SFHs) in these objects. \cite{bhatawdekar_uv_2021} push the redshift boundary even further providing evidence of a MS between $6<z<9$. 

There has also been significant efforts to constrain the lower-mass end of the main sequence. \cite{tasca_evolving_2015} analyze a sample of star-forming galaxies from the VIMOS (VIsible Multi-Object Spectrograph) Ultra-Deep Survey \citep[VUDS][]{le_fevre_vimos_2015} with confirmed spectroscopic redshifts ranging from $0<z<6$. Their results confirm that the main sequence extends to masses as low as $10^7\,\Msun$ for $0.0<z<0.7$. \cite{boogaard_muse_2018} use the deepest MUSE (Multi Unit Spectroscopic Explorer) observations of the Hubble Ultra Deep Field and the Hubble Deep Field South to similarly constrain the low mass end of the MS for redshifts $0.11<z<0.91$. \cite{santini_star_2017} exploit the gravitational lensing of large foreground clusters to probe the main sequence to masses as low as $10^{7.5}\,\Msun$ for $z<4$ and $10^{8.0}\,\Msun$ for $4<z<6$. 

The specific SFR (sSFR) is defined as SFR divided by stellar mass and gives a measure of the current star formation activity compared to the integrated past history. At a fixed mass, sSFR is analogous to the normalization of the main sequence. If the star formation rate closely follows the evolution of the mass accretion rate onto parent halos, the sSFR will be expected to vary with redshift as $\propto(1+z)^{2.25}$ \citep{birnboim_bursting_2007,neistein_constructing_2008,dekel_cold_2009,fakhouri_merger_2010}. The semi-analytic model of \cite{dutton_origin_2010} predicts such evolution in sSFR, as do hydrodynamical simulations \citep[][]{furlong_evolution_2015, donnari_star_2019}. For $z\gtrsim3$, observational studies appear to agree with the predictions \citep{koprowski_reassessment_2014,koprowski_scuba-2_2016,marmol-queralto_evolution_2016,santini_star_2017}.

The recent work of \cite{leja_new_2021} provides a new framework to derive the main sequence from the density of objects in the mass-SFR plane. They fit to objects in the 3D-HST \citep{skelton_3d-hst_2014} and COSMOS-2015 \citep{laigle_cosmos2015_2016} catalogues with a non-parametric star formation history \citep[SFH][]{leja_how_2019}, finding lower normalization of the main sequence by $\sim0.2-0.5$ dex over $0.2<z<3$. This lower normalization resolves a tension between observations and cosmological simulations such as EAGLE \cite{furlong_evolution_2015} and Illustris-TNG \cite{donnari_star_2019}.

\cite{speagle_highly_2014} and \cite{katsianis_high-redshift_2020} have demonstrated how sensitive the determination of the main sequence is to the measurement of SFR, while \cite{leja_new_2021} demonstrates how sensitive it can be to the chosen SFH. The latest SED fitting codes (e.g. \textsc{magphys} \citealt{da_cunha_simple_2008}, \textsc{beagle} \citealt{chevallard_modelling_2016}, \textsc{prospector} \citealt{leja_deriving_2017,johnson_stellar_2021}, \textsc{cigale} \citealt{boquien_cigale_2019, yang_x-cigale_2020}, \textsc{bagpipes} \citealt{carnall_inferring_2018},   \textsc{bayeSED} \citealt{han_decoding_2012, han_bayesed_2014, han_comprehensive_2019}, Dense Basis \citealt{iyer_reconstruction_2017, iyer_nonparametric_2019} and \textsc{ProSpect} \citealt{robotham_prospect_2020})\footnote{see http:\slash\slash www.sedfitting.org\slash Fitting.html for more codes, as well as Pacifici et al. in prep} are able to constrain a variety of physical parameters per galaxy including SFHs, dust attenuation, metallicities and nebular emission, all of which can have a large impact on the derived masses and SFRs. However, the level at which certain properties can be constrained is sensitively dependent on the available data-set, as demonstrated in \citet[hereafter \CL{}]{curtis-lake_modelling_2021}. They find that the emission-line contribution to rest-frame optical broad-band photometry at high redshifts ($z\sim5$ in that study) leads to poorly constrained, biased SFR and stellar mass estimates whereas medium-band filters can significantly improve the constraints. Current data-sets probing high redshifts do not have access to medium-band filters spanning the rest-frame optical. In fact, beyond $z\sim4$, there are only two main filters probing the rest-frame optical; the 3.6 and 4.5 \um{} bands of the \Spitzer\ space telescope.

\begin{figure*}
    \centering
    \includegraphics[width=0.9\textwidth]{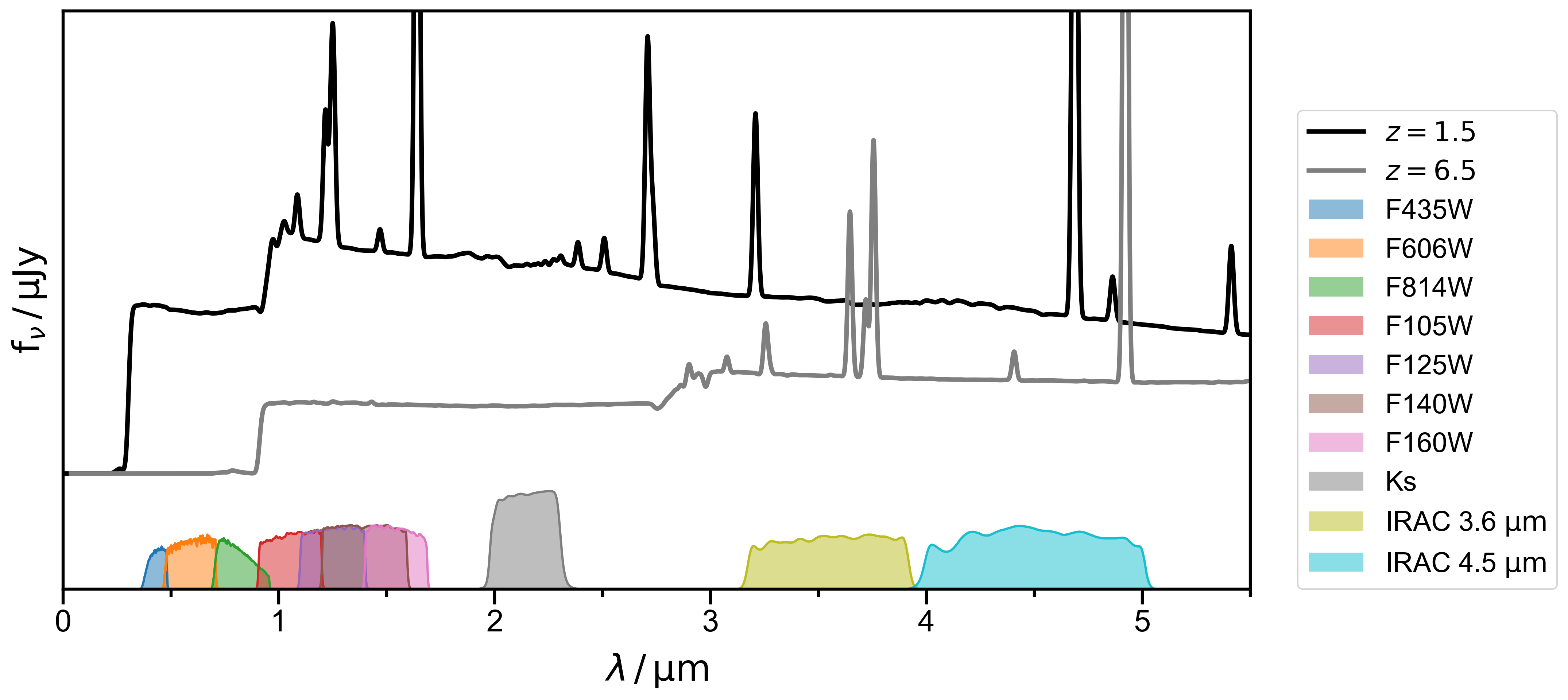}
    \caption{The black and grey lines show example spectra of mock galaxies at redshifts $z=1.5$ and $z=6.5$ respectively. Below the spectra we show the profiles of the ten broad-band filters included in the \astrodeep{} catalogue (see legend). The profiles are plotted with arbitrary normalization and offset from the spectra for clarity. We see that at $z=6.5$ only the IRAC 3.6 and 4.5 \um{} filters sample the rest-frame optical.}
    \label{fig:filter_set}
\end{figure*}

One primary advantage of the latest SED fitting codes is the derivation of robust uncertainties on the derived parameters. However, incorporating these complex, often co-varying uncertainties in population-wide studies requires methods beyond standard linear regression which some studies have been addressing. \cite{kurczynski_evolution_2016} performs sigma-clipping to determine what objects are on the main sequence. They account for co-varying uncertainties by modelling the mass-SFR constraints as single, bivariate Gaussians while fitting to the main sequence in redshift bins. \cite{boogaard_muse_2018} fit a hyperplane in stellar mass, SFR and redshift, self-consistently taking account of the uncertainties using the method of \cite{robotham_hyper-fit_2015}, which models a Gaussian scatter perpendicular to the main sequence. Their sample consists of emission-line selected galaxies from a MUSE survey, so star-forming galaxy selection is based on emission-line properties. \cite{santini_star_2017} and \cite{pearson_main_2018} forward model the main sequence before comparing to observations within redshift bins. \cite{leja_new_2021} use an innovative normalizing flow to measure the density in mass-SFR-redshift, defining the main sequence as the ridge in this space in order to avoid parametrizing the main sequence and outlier distributions separately. They sample from the individual object posterior probability distributions to marginalize over the uncertainties in mass and SFR.

\CL{} suggest a Bayesian hierarchical method to model the main sequence. With this work we extend their approach to include redshift dependence as well as an explicit model to account for outliers. We re-visit the Hubble Frontier Fields, studied by \cite{santini_star_2017}, using the \astrodeep\ catalogues to probe to lower masses and higher redshifts than achievable in blank fields, in order to provide constraints on the low-mass end of the main sequence over a wide redshift range from a consistent data set. We re-visit this data set with self-consistent SFR and mass constraints derived with \beagle\ and fit a fully Bayesian hierarchical, redshift-dependent model of the low-mass, linear portion of the main sequence. We investigate the limitations of this data-set with respect to constraining mass and SFR of individual galaxies with \beagle, demonstrating how to determine when these constraints are robust and how poor constraints can impact the measurements of the main sequence. This data-set represents the best achievable sampling of galaxy SEDs at very high redshifts ($z\gtrsim3$) before we have data from \JWST. In this sense, it provides a representative view of the limitations of what we can measure currently. This study will aid in the understanding of any differing constraints derived with \JWST\ at very high redshifts.

The layout of this paper is as follows: Section~\ref{s:Data and SED fitting with BEAGLE} describes the data and our SED fitting method; Section~\ref{s:Modelling the main sequence} outlines our model of the star-forming main sequence; in Section~\ref{s:Results} we present our results; Section~\ref{s:Discussion} discusses the potential biases and limitations of the data-set for constraining the main sequence, as well as of our method and in Section~\ref{s:Conclusions} we summarize our conclusions. 

Throughout this work we have assumed a \cite{chabrier_galactic_2003} IMF with an upper mass cutoff of 100\,\Msun. We employ a flat $\mathrm{\Lambda CDM}$ cosmology with $\Omega_\Lambda = 0.7$, $\Omega_\mathrm{M} = 0.3$ and $\mathrm{H_0} = 70 \mathrm{\,km \,s^{-1} \,Mpc^{-1}}$. Magnitudes are in the AB system. 

\section{Data and SED fitting}
\label{s:Data and SED fitting with BEAGLE}

The \astrodeep{} catalogue \citep{merlin_astrodeep_2016, castellano_astrodeep_2016, di_criscienzo_astrodeep_2017} includes four of the six Frontier Fields: Abell 2744, MACS0416, MACS0717 and MACS1149, as well as their corresponding parallel fields. The \HST{} Advanced Camera for Surveys (ACS) provides optical imaging while \HST{} Wide-field Camera 3 (WFC3), ground-based HAWK-I (High Acuity Wide field K-band Imager) and \Spitzer{} IRAC (Infrared Array Camera) provide imaging over the near infrared. These provide a total of ten filters that are displayed in Fig.~\ref{fig:filter_set}.

\cite{merlin_astrodeep_2016} and \cite{di_criscienzo_astrodeep_2017} describe how the catalogues are produced but we summarize the main points here. The F160W image is used for primary object detection and provides the base of the catalogue. New objects detected in a stacked IR image (F105W, F125W, F140W and F160W-band) are added to the catalogue. The total \astrodeep{} catalogue contains 29,373 objects. For the purpose of this work we only use the cluster fields, containing 15,379 objects. The HAWK-I Ks-band imaging and \Spitzer{} IRAC imaging has significantly poorer resolution than the \HST{} data. Therefore the \astrodeep{} team use a deconfusion method, using the software \tphot{} \citep{merlin_t-phot_2015, merlin_t-phot_2016}, to perform photometry in these longer wavelength images, taking the high-resolution \HST{} detection image as a prior of the source shapes and positions.

The catalogue includes quality flags that we use to run a first pass selection of objects to analyse. We discarded all objects with $\mathrm{RELFLAG} = 0$ \footnote{This flag value implies unreliable photometry due to either a flagged error from \sextractor{} \citep{bertin_sextractor_1996}, unpyhsical flux in the detection band, less than five reliable \HST{} measurements or close proximity in the image to foreground clusters, stellar spikes or the frame edge.} leaving 11,818 objects.

\subsection{SED fitting}

We wish to exploit the full form of the posterior distribution in stellar mass, \Mstar\ $[=\log(\MstarInLog/\Msun)]$, star formation rate, \sfr\ $[=\log(\sfrInLog/\Msun\,\yr^{-1})]$, and redshift, \redshift\ to derive constraints on the main sequence and its evolution. Although the \astrodeep\ team supplied photometric redshifts and derived physical parameters, to achieve our goal we re-fit to the photometry using \beagle{} (BayEsian Analysis of GaLaxy sEds), a Bayesian SED fitting code \citep{chevallard_modelling_2016}. A detailed description of the \beagle{} parameters which can be adjusted is given in table 2 of \cite{chevallard_modelling_2016}. We do not use the full flexibility of \beagle{} and limit our exploration to the parameters listed in Table~\ref{tab:beagle_priors}, which we describe briefly in this section.

\beagle{} was written to incorporate physically consistent models of nebular plus stellar emission. For this work we model the stellar emission using the version of the \cite{bruzual_stellar_2003} stellar population synthesis models described in \citet[][see their paper for more details]{vidal-garcia_modelling_2017}. For the nebular emission (line and continuum) we adopt the ionization-bounded nebulae models of \cite{gutkin_modelling_2016} that self-consistently trace the production and transmission through the interstellar medium (ISM) of the light from the youngest stars ($<10\,\Myr$).

We characterize the nebular emission using galaxy-wide ionized gas parameters: the interstellar metallicity \Zism{}, which we set equal to the metallicity of the young ionizing stars \Z{}; the typical ionization parameter of a newly ionized \hii{} region, \Us{},\footnote{Note that \Us{} differs from the volume-averaged ionization parameter, $\langle U\rangle$ according to $\langle U\rangle=9/4\,\Us{}$. } which characterizes the ratio of the photon density to hydrogen density at the inner edge of the Str{\"o}mgren sphere; and the mass fraction of interstellar metals in the galaxy locked into dust grains \xid{}. \CL{} demonstrates that \logUs\ and \xid\ are poorly constrained from broad-band photometric data and can bias main sequence determinations. We thus fix \xid\ to 0.3, and impose a relation between \logUs\ and \Zism\ taken from observations:
\begin{equation}
\logUs = -3.638 + 0.055\Z + 0.68\Z^2
\end{equation} 
\noindent
This relation is taken from the observational data presented in \cite{carton_inferring_2017} (priv. communication).
Similarly to \CL{}, within the \hii{} region we fix the carbon to oxygen abundance ratio \CO{} to the solar value of $\COsun=0.44$, the hydrogen density to $\nh=100 \mathrm{\,cm^{-3}}$ and model the intergalactic absorption as prescribed by \cite{inoue_updated_2014}.

To maintain consistency with previous observational studies \citep[e.g.][]{kurczynski_evolution_2016, santini_star_2017}, we adopt a delayed exponentially declining (DE) SFH of the form $\sfrInLog(\msa) \propto \msa \, \exp(-\msa/\tausfr)$ where $\sfrInLog(\msa)$ is the star formation rate, $\msa$ is the time since the formation of the oldest stars and $\tausfr$ is the time between the onset of star formation and the peak of the SFH. This SFH allows for very low SFR at a given stellar mass (not allowed by a constant SFH), while also describing a rising SFH when $\msa<\tausfr$, which has been suggested to be suitable for high redshifts \citep[e.g.][]{salmon_relation_2015}. The integral of the SFH with respect to time gives the total amount of stellar mass formed, \MtotInLog{}, and is the parameter sampled over within \beagle. \MstarInLog{} gives the stellar mass in stars at a given time after accounting for the return fraction to the ISM after stars die, and is the parameter used for our measurements of the main sequence.

\CL{} show that fitting to \JWST{} broad-band fluxes of $z\sim5$ simulated galaxies with a DE SFH results in poorly constrained physical parameters which in turn biases the measurement of the main sequence. This is due to the unknown contribution of emission line fluxes to the broad-band filters, and the effects are mitigated when medium-band filters are available. Our data-set does not include medium-band filters so where emission lines contribute a significant fraction of the broad-band flux at high redshifts, we may still derive biased stellar masses and SFRs. At lower redshifts, however, the line equivalent widths are lower and hence the relative contribution of emission lines compared to the stellar continuum is much smaller. We investigate the effect of poor constraints on our derived main sequence parameters in Section \ref{ss:Choice of Star Formation History}.

\begin{table}[ht]
    \centering
    \caption{Parameters and associated priors set in \beagle{} for fitting to the \astrodeep{} catalogue.}
    \label{tab:beagle_priors}
    \begin{tabular}{c c}
        \toprule
        Parameter & Prior \\
        \midrule
        $\logmsa$ & $\mathcal{N}(8.0, 2.0^2), \mathrm{truncated} \in [6.0, 10.0]$ \\ 
        $\logtausfr$ & $\mathrm{Uniform} \in [7.0, 10.5]$ \\ 
        $\log(\MtotInLog/\Msun)$ & $\mathrm{Uniform} \in [5.0, 12.0]$ \\ %
        $\log(\Z/\Zsun)$ & $\mathrm{Uniform} \in [-2.1, 0.3]$ \\ 
        $\redshift$ & $\mathrm{Uniform} \in [0.0, 15.0]$ \\ 
        $\tauV$ & $\exp(-\tauV)$, for $\tauV \in [0.0, 6.0]$ \\ 
        $\logUs$ & Dependent \\ 
        $\xid$ & Fixed 0.3 \\ 
        $\mud$ & Fixed 0.4 \\ 
        $\nh/\cm^{-3}$ & Fixed 100 \\ 
        $\CO/\COsun$ & Fixed to solar, where $\COsun = 0.44$ \\ 
        $\Mup/\Msun$ & Fixed 100 \\ 
        \bottomrule
    \end{tabular}
\end{table}

We incorporate dust attenuation using the physically motivated two-component model of \cite{charlot_simple_2000}. The components of this model are the diffuse dust distributed uniformly throughout a galaxy's ISM, and the dust within denser stellar birth clouds. Within this model, stars older than $10\,\Myr$ only see the effects of diffuse dust within the ISM, having a \textit{V}-band optical depth equal to that of the ISM, \tauVism{}. The birth clouds enshrouding stars younger than $10\,\Myr$ have an optical depth \tauVbc, giving a total optical depth to young stars of \tauV = \tauVism + \tauVbc. The fractional attenuation of stars residing in the ISM compared to those residing in stellar birth clouds is given by:
\begin{equation}
\mud = \frac{\tauVism}{\tauVism + \tauVbc}
\end{equation}
\noindent
We use the updated treatment of dust in \beagle{}\footnote{Available from \beagle{} v0.27.1.} which accounts for the effects of dust within the \cite{gutkin_modelling_2016} nebular models themselves, as described in \CL{}, section 2.

\begin{figure*}
    \centering
    \includegraphics[width=0.9\textwidth]{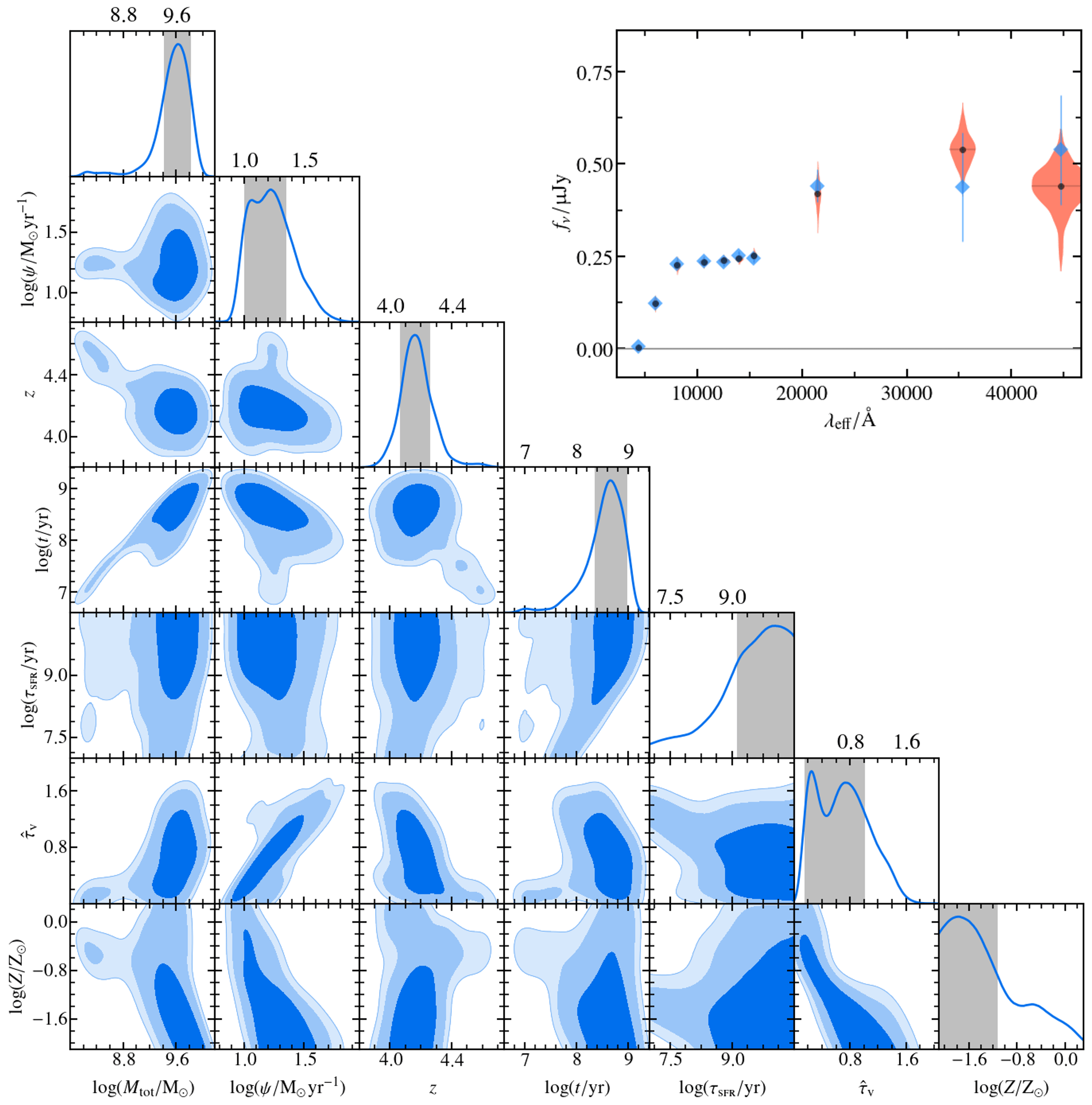}
    \caption{Abell 2744 cluster, ID 331. \textit{Bottom left:} The diagonal panels show the marginal probability distributions for each of the six fitted parameters ($\log(\MtotInLog/\Msun)$, \redshift{}, \logmsa{}, \logtausfr{}, \tauV{} and $\log(\Z/\Zsun)$) as well as $\log(\sfrInLog/\Msun\,\yr^{-1})$. The other panels show the joint posterior distributions for every pair of parameters. \textit{Top right:} Blue diamonds represent the observed SED. Orange violins show the predicted model fluxes as determined by the posterior probability distributions of the fitted parameters.}
    \label{fig:304_triangle_marginal}
\end{figure*}

We fit to 11,818 objects within the four cluster fields, with six free parameters: (\logmsa{}, \logtausfr{}, \MtotInLog{}, \Z{}, \redshift{} and \tauV{}). Table \ref{tab:beagle_priors} shows the prior distributions configured within \beagle{}. 

We do not include HAWK-I and \Spitzer{} photometry in the fitting if the \astrodeep{} COVMAX flag indicates that an object suffers severe blending with another source during the \tphot{} extraction process ($\mathrm{COVMAX}_{\mathrm{filt}}>1$). The COVMAX flag is publicly available for the Abell 2744, MACS0416 and their parallels while the \astrodeep{} team provided the flags for MACS0717 and MACS1149 (priv. communication).

When fitting to the observed photometry, we allow for a minimum relative error which is added in quadrature to the measurement uncertainties. This minimum error accounts for the possible calibration differences from photometry derived with different telescopes, as well as the uncertainties in the models. For the \HST\ photometry, we allow a minimum relative error of 0.04. This is higher than previously suggested \citep[e.g.][]{chevallard_modelling_2016} after finding that the brightest galaxies had poor-fitting $\chi^2$ values due to the small measurement uncertainties. Values of 0.05 and 0.1 are applied to the HAWK-I and IRAC photometry respectively. HAWK-I and IRAC images require deconfusion, and hence likely suffer systematic uncertainties that are not accounted for in the supplied photometric errors. 

Fig.~\ref{fig:304_triangle_marginal} is an example (Abell 2744 cluster, ID 331) of the \beagle{} output available for each fitted object.

\subsection{Photometric redshift analysis}

The \astrodeep{} collaboration provide photometric redshift estimates which are the median values taken from six independent methods as described in \cite{castellano_astrodeep_2016} (Abell 2744 and MACS0416) and \cite{di_criscienzo_astrodeep_2017} (MACS0717 and MACS1149).

Prior to analysing \beagle{}-derived photometric redshifts, we discard objects with a F160W AB magnitude fainter than 27.5. This cut was employed by \cite{santini_star_2017}, and based on simulations by \cite{merlin_astrodeep_2016} designed to determine the detection completeness of the images. The limit corresponds to 90 - 95\% completeness for point-like sources and 50 - 80\% for extended disks with a 0.2" half-light radius. In addition, we reject objects with a poor fit by \beagle{} defined as having a minimum $\chi^2>13.28$.\footnote{Fits with a minimum $\chi^2 = 13.28$ are consistent with our model 99\% of the time, under the assumption of ten available \astrodeep{} filters, with \beagle{} fitting for six independent parameters.} We also impose a lower mass cut as described in Section~\ref{ss:Sample selection for main sequence analysis}.

\begin{figure}
    \centering
    \includegraphics[width=0.45\textwidth]{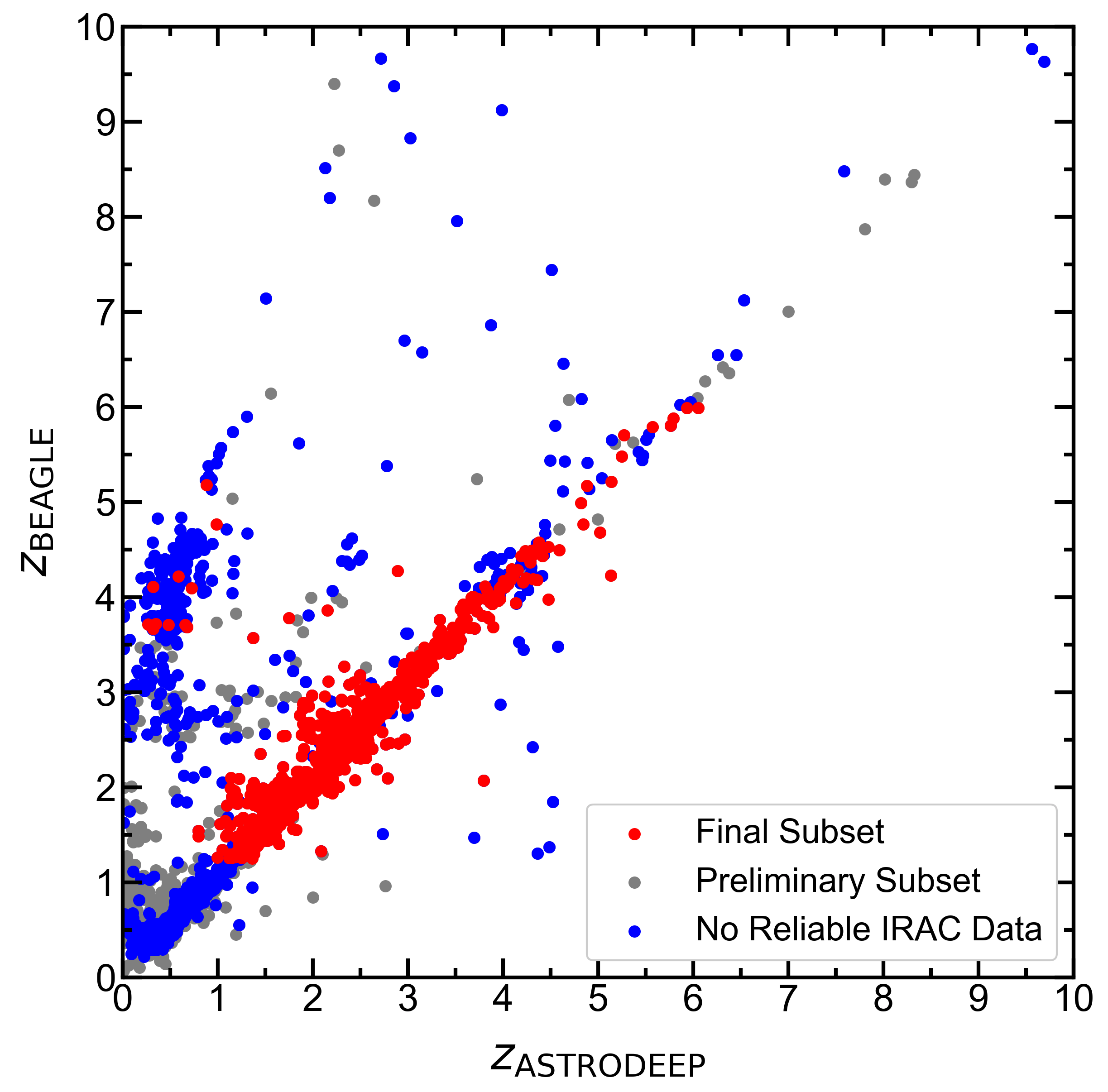}
    \caption{\beagle{}-derived photometric redshifts (posterior medians) plotted against \astrodeep{} redshifts. All objects with $\mathrm{RELFLAG}=1$, F160W magnitude $<27.5$, \beagle{}-fitted minimum $\chi^2<13.28$ and a redshift-dependent lower mass cut applied prior to correcting for gravitational lensing (see text and Fig.~\ref{fig:lower_mass_limit}) are plotted.  The red points mark the \totsample\ galaxies chosen as our final subset (see text). Blue points show the objects which have no reliable IRAC data and that do not make it into our sample. Grey points have good IRAC photometry but do not make it into our sample.}
    \label{fig:redshift_vs_redshift}
\end{figure}

In Fig.~\ref{fig:redshift_vs_redshift} we compare the \beagle{}-derived (posterior median) photometric redshifts to those in the \astrodeep{} catalogue \citep[see][section 3]{castellano_astrodeep_2016}. The plot shows objects from the four cluster fields which satisfy the above criteria. Whilst the majority of objects lie close to the identity relation, there are many which \beagle{} has identified as $z \sim 4$ in contrast to an \astrodeep{} redshift of $z \sim 0.5$. Photometric redshifts are primarily determined by the detection of a break in the observed SED. In this scenario \astrodeep{} has assigned a Balmer break (at rest-frame 3646\A{}) to the observed break while \beagle{} has assigned a Lyman break (at rest-frame 1216\A{}). 

For this filter-set the Lyman break is not reliably bracketed by two filters until $z \sim 4.5$. At redshifts lower than this, reliable determination of Lyman vs. Balmer break will be improved with the Ks and IRAC bands sampling red-ward of the Balmer break. The majority of objects with disagreement between \beagle{} and \astrodeep{} lack robust IRAC photometry (as shown by blue points on the plot), and therefore only show one observed break in the SED. In this situation, for any given object and photometric redshift code, there is some probability that the observed break is incorrectly assigned. For different codes this probability will vary depending on template set and priors. \cite{castellano_astrodeep_2016} takes the median value of multiple (six) codes, thus mitigating this issue if at least 50\% of the codes choose the correct value. Since we require rest-frame optical photometry for our \Mstar{} constraints, those objects with poor photometric redshift estimates would be rejected at the next stage, even if \beagle{} \textit{had} agreed with the \astrodeep{} determinations.

By $z\sim4.5$, the Balmer break falls red-wards of the Ks-band. We therefore require objects above $z>3.5$ to have at least one robust photometric point from the three longest wavelength filters (Ks, 3.6 \um{}, 4.5 \um{}), while above $z>4.5$, we require at least one IRAC flux point. Those objects that lack good IRAC/Ks photometry at these redshifts tend to be due to significant confusion in the \Spitzer{} images. This is not dependent on the intrinsic properties of the objects themselves, rather the projected distribution of sources on the sky. We therefore do not expect this cut to significantly bias our main sequence determination. Furthermore, we apply a lower redshift limit of 1.25 as below this the F435W-band no longer probes the rest-frame far ultra-violet required for secure SFR determination.

We visually inspected the images and SEDs for all objects with either a \beagle{} redshift ($z_{\beagle{}}$) or an \astrodeep{} redshift ($z_{\astrodeep{}}$) of greater than 3.5. We leverage the better accuracy of the \astrodeep{} photometric redshifts by discarding remaining objects (with both $z_{\beagle{}}$ and $z_{\astrodeep{}}<3.5$) if $|z_{\beagle{}} - z_{\astrodeep{}}|>1$.

\subsection{Sample selection for main sequence analysis}
\label{ss:Sample selection for main sequence analysis}

\begin{figure}
    \centering
    \includegraphics[width=0.45\textwidth]{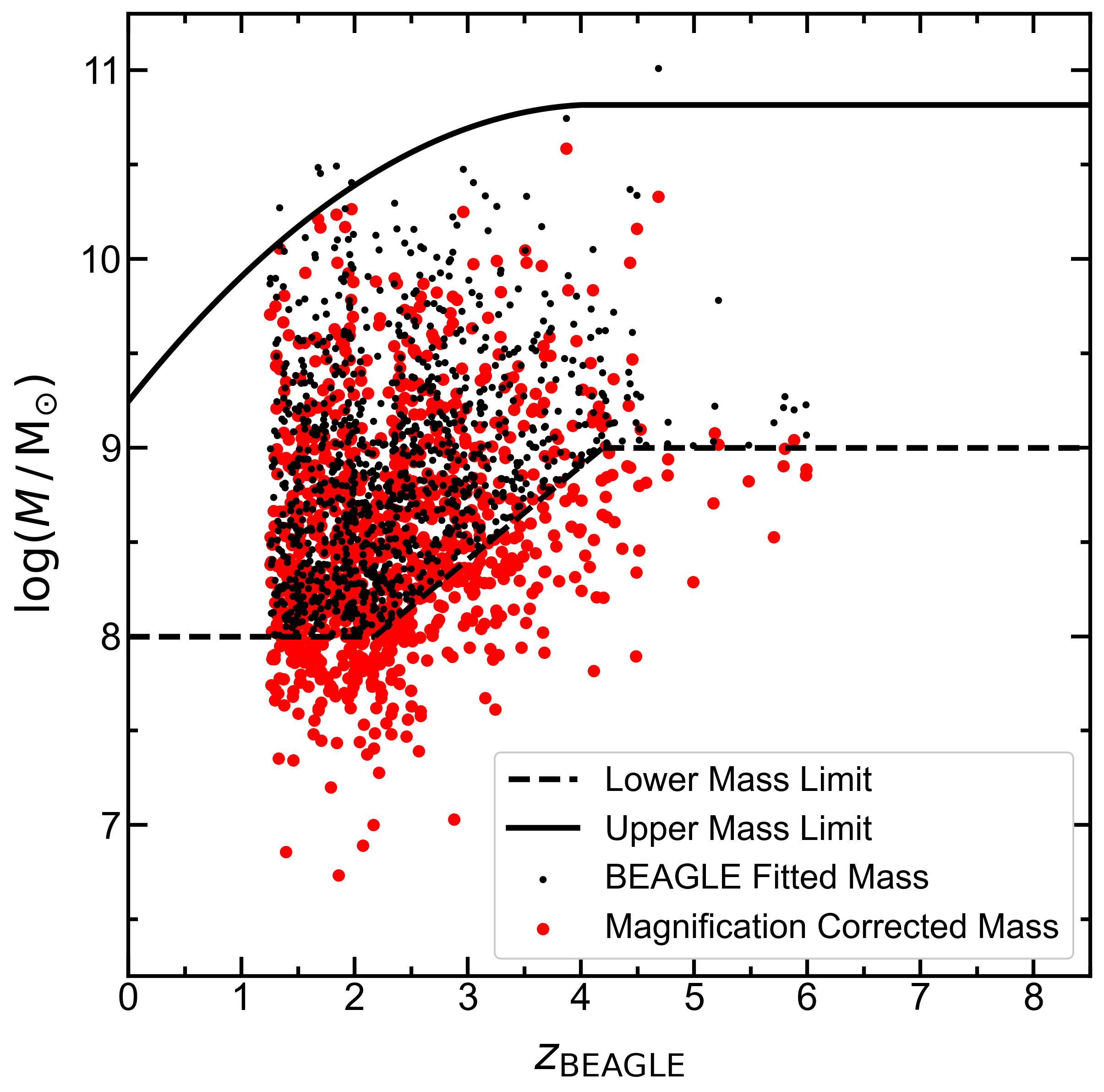}
    \caption{Shown in blue are the \beagle{}-derived posterior median stellar mass and redshift estimates plotted against each other for our final sample of \totsample\ objects. The dashed black line shows the lower limit imposed upon the \beagle{}-derived stellar masses based on 95\% mass completeness for F160W magnitude $<27.5$ (see text for details). The cuts are imposed prior to correcting the derived properties for the effects of gravitational lensing. Magnification-corrected stellar masses are shown in red. The solid black line shows the redshift-dependent turnover mass as fitted by \protect\cite{tomczak_sfr-m_2016}, fixed as a constant for $z>4$.  This upper limit is applied after magnification corrections.}
    \label{fig:lower_mass_limit}
\end{figure}

For analysing the main sequence we need a sample that is complete in stellar mass.  We therefore impose a redshift-dependent mass cut in our samples. This mass limit was calculated using the JAGUAR mock catalogue \citep{williams_jwst_2018}, which was produced with the same stellar and nebular models, making the limits self-consistent with the \beagle\ fits performed here. We calculate the mass limit, as a function of redshift, at which the sample is 95\% complete in stellar mass for F160W magnitude $<27.5$. The limit is displayed as the dashed black line in  Fig.~\ref{fig:lower_mass_limit}. The limit is a function of the position of the main sequence in the \MstarSFR\ plane, how well the F160W limit approximates a stellar mass limit, and how the brightness of the objects vary with redshift.  At low redshifts the F160W cut approximates a stellar mass cut, whereas at high redshifts it approximates a cut in SFR, where the transition between these two limits causes an increase in the lower mass limit between $z\sim2.5-4$.  Where the mass limit is approximately flat, the change in position of the objects in the \MstarSFR\ plane must be compensating the reduction in flux with increasing redshift.  We apply the cuts based on \Mstar\ estimates prior to correcting them for the effects of gravitational lensing (as the F160W is a limit of the image, not the intrinsic galactic properties). These are shown as blue points in Fig.~\ref{fig:lower_mass_limit}.  The red points show the masses after lensing is accounted for, demonstrating that we probe below the \Mstar\ limits of standard blank fields. We correct the \beagle-derived masses and SFRs using the magnification value supplied in the \astrodeep\ catalogues \citep[see][for details]{castellano_astrodeep_2016}. 

A redshift-dependent upper mass limit is also imposed on the magnification-corrected values to ensure that we are not including objects in the regime where the main sequence has been observed to flatten. Between $0<z<4$ we take the parametrization of the turnover mass from \cite{tomczak_sfr-m_2016} and for $z>4$ we choose a fixed turnover mass of $\sim10^{10.8}\,\Msun$. This limit is shown as the thick black line in Fig.~\ref{fig:lower_mass_limit}. 

In summary, the full set of selection criteria includes selecting objects with reliable photometry identified by $\mathrm{RELFLAG} = 1$ with F160W magnitude $<27.5$ and \beagle\ fits with $\chi^2<13.28$.  We require agreement with \astrodeep{} redshift within $|\Delta\redshift|<1$ for $\redshift<3.5$ and visual inspection above $\redshift>3.5$. We ensure objects have photometry sampling the rest-frame optical, enabling stellar mass determination.  Finally we apply the upper and lower mass limits described here.  Our final sample spans $1.25<z<6$ and includes \totsample\ objects which are shown as red points in Fig.~\ref{fig:redshift_vs_redshift} and Fig.~\ref{fig:lower_mass_limit}.

\section{Modelling the main sequence}
\label{s:Modelling the main sequence}

In this section we detail the steps that we have taken to model the star-forming main sequence spanning redshifts $1.25~<~z~<~6$.

At a single redshift, ordinary linear regression applied to the star-forming main sequence fails to fully account for heteroskedastic, co-varying errors and the non-uniform distribution of \Mstar{}. \citet[hereafter \kelly{}]{kelly_aspects_2007} proposes a Bayesian hierarchical method to address these concerns, which has been extended by \CL{} to work with the output joint posteriors of \Mstar{} and \sfr{} derived from SED fitting with \beagle{}. This approach allows for the self-consistent propagation of measurement uncertainties onto the parameters which describe the main sequence relation: the slope, intercept and intrinsic scatter. 
 
Throughout this section we refer to Bayesian terms such as prior probability, likelihood and posterior probability. It is therefore informative to recap Bayes' theorem, which states that the posterior probability distribution of the model parameters, $\conditional{\ThetaB}{\DB,H}$, can be expressed as:
\begin{equation}
    \conditional{\ThetaB}{\DB,H} \propto \conditional{\DB}{\ThetaB,H} \, \conditional{\ThetaB}{H}
\end{equation}
\noindent
where $\conditional{\DB}{\ThetaB,H}$ is the likelihood of the data $\DB$ given a model (or hypothesis), $H$, with associated parameters, $\ThetaB$. The prior probability, $\conditional{\ThetaB}{H}$, describes the knowledge we have of the model before analysis of the data.

\CL{} apply their model to a mock photometric sample of main sequence galaxies at $z \sim 5$. They model the main sequence as a linear relation with Gaussian scatter, which we re-write, subtly, to find the normalization of the relation at $\log(\MstarInLog/\Msun)=9.7$:
\begin{equation}
    \sfr = \intercept + \slope(\Mstar-9.7) + \mathcal{N}(0,\scatter^2)
\label{eqn:main_sequence}
\end{equation}
\noindent
\intercept\ is the normalization of the main sequence at a stellar mass of $\log(\MstarInLog/\Msun)=9.7$, \slope\ is the slope, and $\mathcal{N}(0,\scatter^2)$ denotes a Gaussian distribution centred on zero with a variance of $\scatter^2$ and describes the intrinsic scatter about the relation. Throughout this paper, when describing SFR and stellar mass in log space, we use \sfr\ $[=\log(\sfrInLog/\Msun\,\yr^{-1})]$ and \Mstar\ $[=\log(\MstarInLog/\Msun)]$.

The three levels of the K07 Bayesian hierarchical model are: the distribution of stellar masses, which is not assumed to be uniform; the distribution of \sfr\ given \Mstar (equation~\ref{eqn:main_sequence}); and the lowest level describes data given the unknown true \Mstar\ and \sfr\ values. In our case the data consists of photometric fluxes and uncertainties (see CL21, section 3.4 for more details). The K07 model is designed to marginalize over the unknown, true values of \Mstar\ and \sfr\ for each object when deriving the parameters of interest, namely \intercept, \slope\ and \scatter. 

In this work, we extend the model of CL21 by including the redshift evolution of the main sequence. It is also important to account for objects which do not belong to the star-forming main sequence, which we shall refer to as ``outliers''. We explicitly model these outliers to ensure that the uncertainty of which objects belong to the main sequence is fully accounted for in our analysis. To determine what form of redshift evolution to include in the model, we first measure the main sequence in a series of redshift bins (Section~\ref{ss:Redshift Bins}). We describe our model for the redshift evolution of the main sequence in Section \ref{ss:Redshift Evolution of the MS}. 

\subsection{Redshift bins}
\label{ss:Redshift Bins}

\subsubsection{Modelling outliers}
\label{sss:Modelling Outliers}

\begin{figure*}
    \centering
    \includegraphics[width=0.99\textwidth]{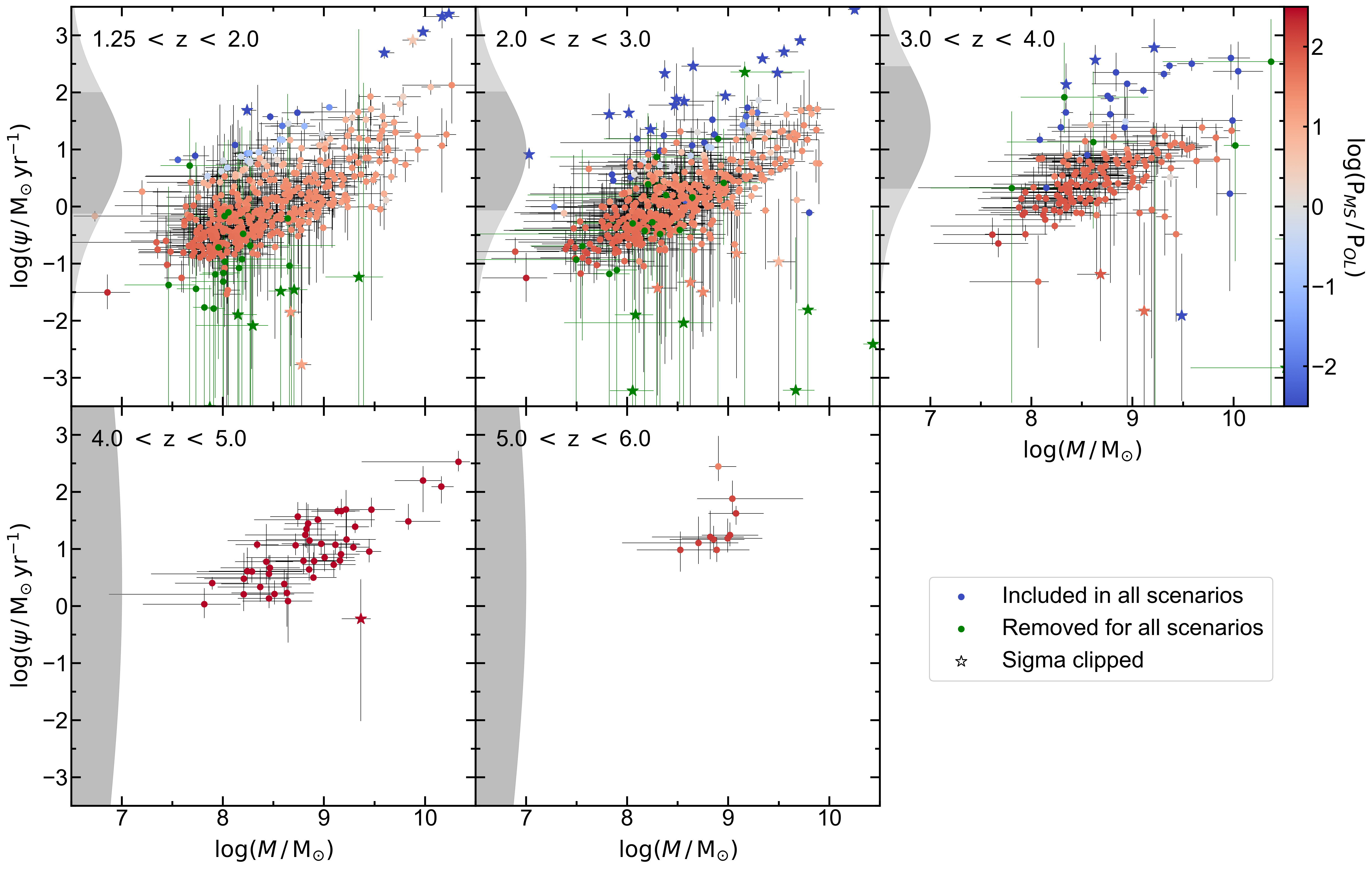}
    \caption{\beagle{}-derived posterior median $\log(\sfrInLog{}/\Msun\,\yr^{-1})$ plotted against $\log(\MstarInLog{}/\Msun)$ in redshift bins. The error bars show marginalized 68\% credible intervals in these two parameters. The red-blue colour-coding represents logarithm of the ratio of probability that a given object belongs on the main sequence to the probability that it is an outlier (see equation~\ref{eqn:outlier_model}). Green symbols represent the objects removed, regardless of outlier treatment, due to poorly constrained \sfr{}. Stars of any colour show the objects removed during the $3\sigma$-clipping procedure for the \OLclipped{} method. The grey histograms on the left of each panel represent the best-fitting outlier distribution, showing 68\% credible regions with dark grey. In the bottom two panels the outlier distribution is shown as broad, since the distribution is unconstrained due to lack of obvious outliers in redshift bins $4<z<5$ and $5<z<6$.}
    \label{fig:redshift_bins_MS}
\end{figure*}

Not all galaxies belong to the star-forming main sequence. Quiescent galaxies will lie significantly below the main sequence while star-bursting galaxies, which may be experiencing a recent or ongoing merger, can lie significantly above the relation. In order to investigate how to appropriately model the outliers in our sample, we initially divide our subset of \totsample\ objects based upon their \beagle{}-derived posterior medians into redshift bins of $1.25<z<2$, $2<z<3$, $3<z<4$, $4<z<5$ and $5<z<6$. 

\cite{hogg_data_2010} suggest a simple model for incorporating outliers. We therefore investigate the possibility of extending the work of \CL{} using this model which allows objects to either reside on the main sequence or within a separate outlier distribution which is described as a simple Gaussian:
\begin{equation}
    \sfr \sim \mathcal{N}(\OLmean, \OLscatter^2),
\label{eqn:OL}
\end{equation}
\noindent
where $\mathcal{N}(\OLmean, \OLscatter^2)$ is a normal distribution with mean \OLmean{} and standard deviation \OLscatter{}.   This model is implemented as follows:
\begin{equation}
\begin{aligned}
    \conditional{\sfr}{\Mstar} &=  \PMS + \POL\\
    \PMS &= (1-\OLprob)\,\conditional{\sfr}{\intercept + \slope(\Mstar-9.7) + \mathcal{N}(0,\scatter^2)}\\
    \POL &= \OLprob\,\,\conditional{\sfr}{\mathcal{N}(\OLmean, \OLscatter^2)}\\
\end{aligned}
\label{eqn:outlier_model}
\end{equation}
\noindent
where \PMS\ is the probability that the object belongs to the main sequence and \POL\ is the probability that the object is an outlier. The parameter \OLprob{} defines the ratio of the integrals of the functions describing the outlier and main-sequence distributions at \Mstar, respectively. We restrict $\OLprob<0.5$, and $\OLscatter{}>1$ ensuring that within the main sequence the relative probability of any given object being an outlier is very small. However, where the probability that an object belongs to the main sequence becomes negligible, there is higher probability that the object belongs to the outlier distribution. When implementing this outlier model within redshift bins we have to make a decision about how we treat the mass distribution. We make the assumption that the distribution of \Mstar{} in the outlier population is the same as that of the objects on the main sequence.

\begin{figure*}
    \centering
    \subfigure{\includegraphics[width=0.45\textwidth]{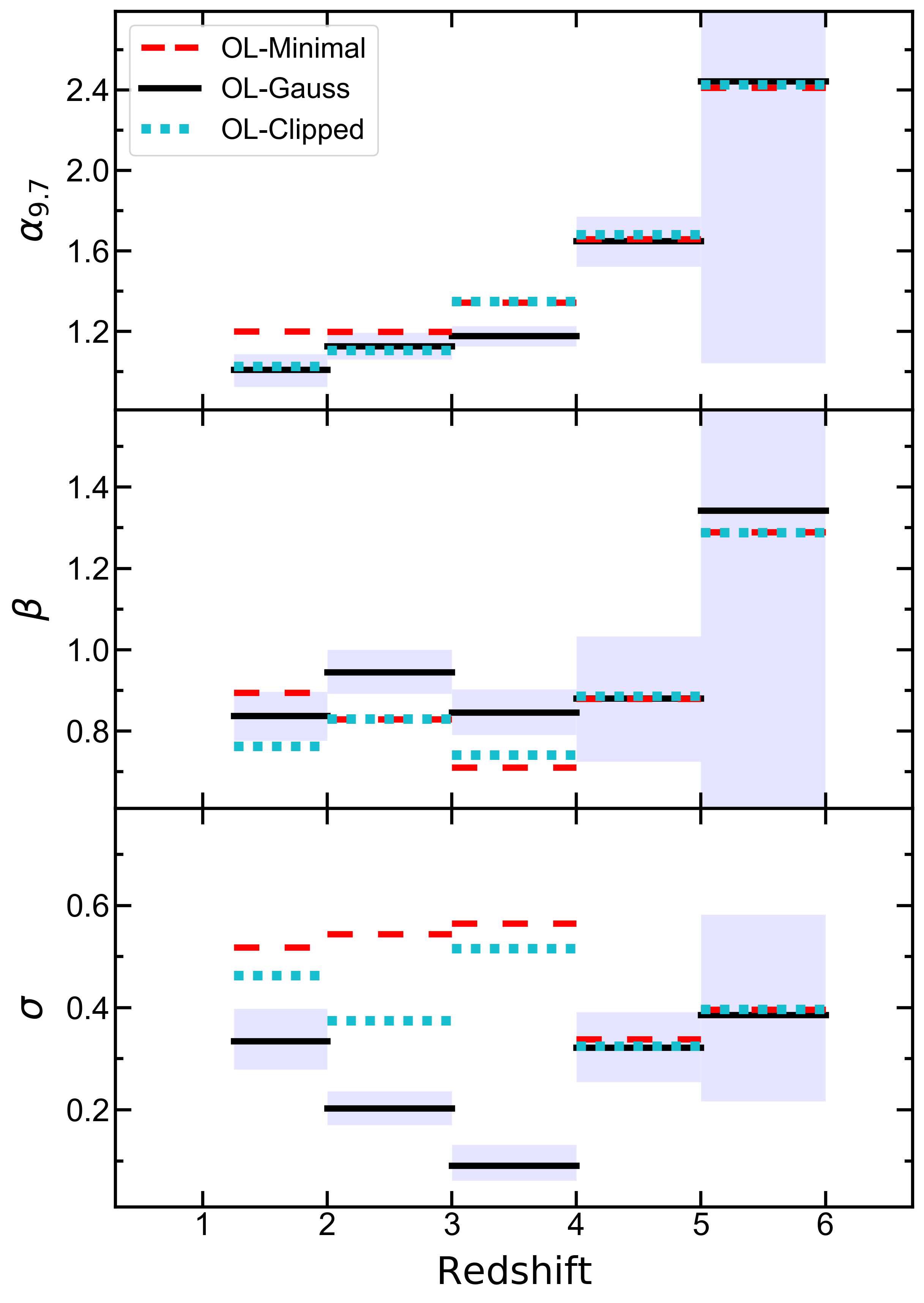}}
    \subfigure{\includegraphics[width=0.45\textwidth]{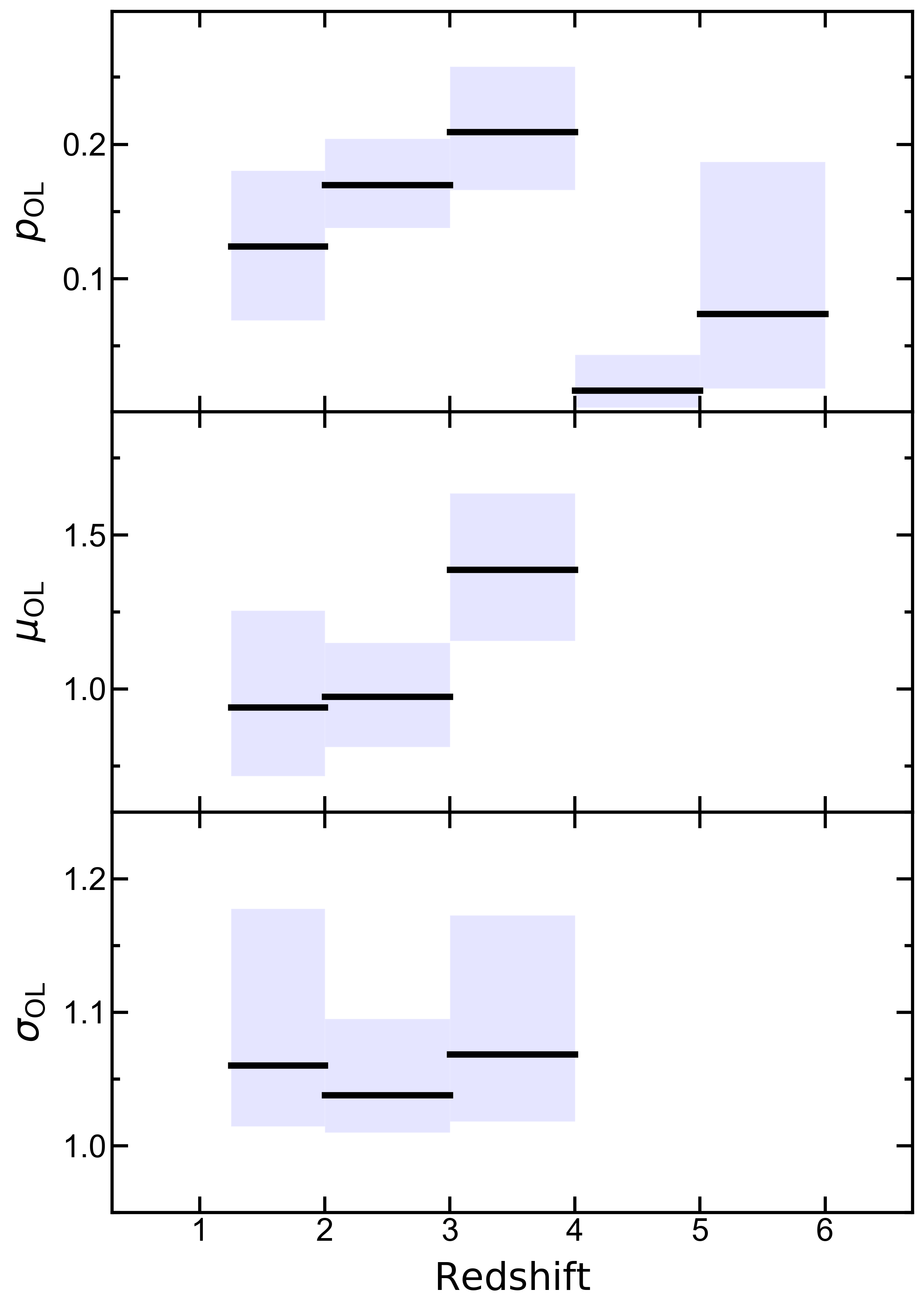}}
    \caption{Redshift bin results showing the posterior median values for the best fit \OLnone{}, \OLhogg{} and \OLclipped{} methods (as shown in the legend). The shaded blue regions show the 68\% credible intervals for the \OLhogg{} method. \textit{Left panel:} From top to bottom shows the normalization, slope and intrinsic scatter of the main sequence as a function of redshift. \textit{Right panel:} From top to bottom shows \OLprob{}, \OLmean{} and \OLscatter{} of the outlier distribution as a function of redshift. The redshift bin $4<z<5$ and $5<z<6$ results have been omitted from the bottom two panels for clarity, as in this scenario, \OLprob{} approaches 0.}
    \label{fig:redshift_bins}
\end{figure*}

\begin{table*}
    \centering
    \caption{Posterior median values and 68\% credible intervals for the fitted main sequence and outlier parameters per redshift bin.}
    \label{tab:bin_results}
    \begin{tabular}{c c c c c c}
        \toprule
        Parameter & $1.25<z<2$ & $2<z<3$ & $3<z<4$ & $4<z<5$ & $5<z<6$ \\
        \midrule
        \intercept & $1.01^{+0.08}_{-0.09}$  & $1.13^{+0.07}_{-0.07}$  & $1.18^{+0.05}_{-0.05}$  & $1.65^{+0.12}_{-0.13}$  & $2.44^{+1.51}_{-1.40}$ \\
        \slope & $0.84^{+0.06}_{-0.06}$  & $0.94^{+0.06}_{-0.05}$  & $0.85^{+0.06}_{-0.06}$  & $0.88^{+0.15}_{-0.16}$  & $1.34^{+2.09}_{-1.82}$ \\
        \scatter & $0.33^{+0.06}_{-0.06}$  & $0.20^{+0.03}_{-0.03}$  & $0.09^{+0.04}_{-0.03}$  & $0.32^{+0.07}_{-0.07}$  & $0.39^{+0.20}_{-0.17}$ \\
        \OLprob & $0.12^{+0.06}_{-0.06}$  & $0.17^{+0.03}_{-0.03}$  & $0.21^{+0.05}_{-0.04}$  & $0.02^{+0.03}_{-0.01}$  & $0.07^{+0.11}_{-0.06}$ \\
        \OLmean & $0.94^{+0.31}_{-0.22}$  & $0.97^{+0.17}_{-0.16}$  & $1.39^{+0.25}_{-0.23}$  & $0.35^{+6.39}_{-6.73}$  & $0.52^{+5.94}_{-6.85}$ \\
        \OLscatter & $1.06^{+0.12}_{-0.05}$  & $1.04^{+0.06}_{-0.03}$  & $1.07^{+0.10}_{-0.05}$  & $5.35^{+3.06}_{-2.98}$  & $5.26^{+3.19}_{-2.98}$ \\
        \bottomrule
    \end{tabular}
\end{table*}

During our preliminary tests, it became clear that the majority of the quiescent outliers sitting below the main sequence were highly unconstrained in \sfr{}. This freedom allowed these objects to be modelled as belonging to the main sequence, effectively lowering the measured normalization, biasing the slope and increasing the intrinsic scatter. We decided to remove such poorly constrained outliers beforehand by sampling from each object's posterior distributions of \Mstar{}, \sfr{} and \redshift{}, rejecting objects with standard deviation in $\sfr>2$ for the samples within the redshift bin. This method removed 38, 26, 8, 0 and 0 outliers below the main sequence from the bins $1.25<z<2$, $2<z<3$, $3<z<4$, $4<z<5$ and $5<z<6$ respectively.

Having accounted for the majority of the quiescent outliers below the main sequence, we test the proposed outlier model (we label this method \OLhogg) and compare it to two other methods. The first of these methods calculates the main sequence without any outlier rejection beyond the objects with poorly constrained \sfr{} (we label this method \OLnone).
The second method identifies outliers from iterative $3\sigma$-clipping, where we iteratively remove objects further away than 3 standard deviations from the best linear fit to posterior medians in \Mstar\ and \sfr. This is implemented prior to the removal of the poorly constrained objects below the main sequence. The method of clipping outliers prior to fitting the main sequence is comparable to approaches within the literature \citep[e.g.][]{santini_star_2017, kurczynski_evolution_2016}. We label this method \OLclipped. 

Once outliers are removed for the \OLnone\ and \OLclipped\ methods, the main sequence is measured from the remaining objects using the \CL{} Bayesian hierarchical model.  The \OLhogg\ method measures the main sequence using an updated version of the model, adapted to include the outlier model described in equation~\ref{eqn:outlier_model}.

\subsubsection{Redshift bin results}
\label{sss:redshift_bin_results}

Fig.~\ref{fig:redshift_bins_MS} displays \sfr\ vs. \Mstar\ for the objects in each of the five redshift bins. The points, with various symbols, display the \beagle-derived posterior medians in \Mstar\ and \sfr, while the errors show the marginalized 68\% credible intervals. Objects that are removed because they have poor constraints in \sfr\ are coloured green and objects that are removed by the \OLclipped\ method are displayed as stars. The remaining points are coloured by $\log(\PMS/\POL)$, therefore showing the relative probability of being on the main sequence or within the outlier distribution when using the \OLhogg\ method.

The left three panels of Fig.~\ref{fig:redshift_bins} show the derived posterior median values of the main sequence parameters, \intercept, \slope, and \scatter, in redshift bins spanning $1.25<z<6$. The derived values are also reported in Table~\ref{tab:bin_results}. The blue shaded rectangles and solid lines show the 68\% credible regions and posterior medians, respectively, for the constraints derived with the \OLhogg\ method. The dashed and dotted lines show the posterior medians for the parameters derived with the \OLnone\ and \OLclipped\ methods, respectively.

Fig.~\ref{fig:redshift_bins} (top left panel) shows that all methods measure an increasing normalization with redshift. For the lowest two redshift bins, the \OLnone\ method returns a higher normalization ($\sim 1.2 - 1.3$) than the other two methods ($\sim 1.0 - 1.2$). This is to be expected as the \OLclipped\ and \OLhogg\ methods both identify a fraction of the objects above the main sequence as outliers, lowering the measured normalization. Within the $3<z<4$ bin, however, very few objects are rejected above the relation with the \OLclipped\ method, making the results of the \OLnone\ and \OLclipped\ methods very similar. The \OLhogg\ method, however, ends up assigning a lot of the objects above the relation a high probability of belonging to the outlier distribution, returning a lower normalization. This demonstrates how sensitive the results are to the chosen method to account for outliers. 

Fig.~\ref{fig:redshift_bins} (middle left panel) shows that overall there is no strong evidence of varying slope with any of the three methods. \OLhogg\ measures a steeper slope than the \OLclipped\ and \OLnone\ methods in the $2<z<3$ and $3<z<4$ redshift bins, because some objects slightly above the relation at masses $\Mstar\sim8-9$ have non-zero probability of belonging to the \OLhogg\ outlier distribution (Fig.~\ref{fig:redshift_bins_MS}, top middle and top right panels). In the lowest redshift bin, however, the \OLclipped\ method measures the shallowest slope ($\sim0.76$), but \OLnone\ measured the steepest slope ($\sim0.89$). This is because the \OLnone\ run includes two objects below the relation at $\Mstar\sim8.5-8.7$ that are identified as having poor constraints on \sfr, yet are clearly below the main sequence (Fig.~\ref{fig:redshift_bins_MS}, top left panel).The steeper slope from the \OLnone\ method is therefore less reliable. 

\begin{figure}
    \centering
    \includegraphics[width=0.45\textwidth]{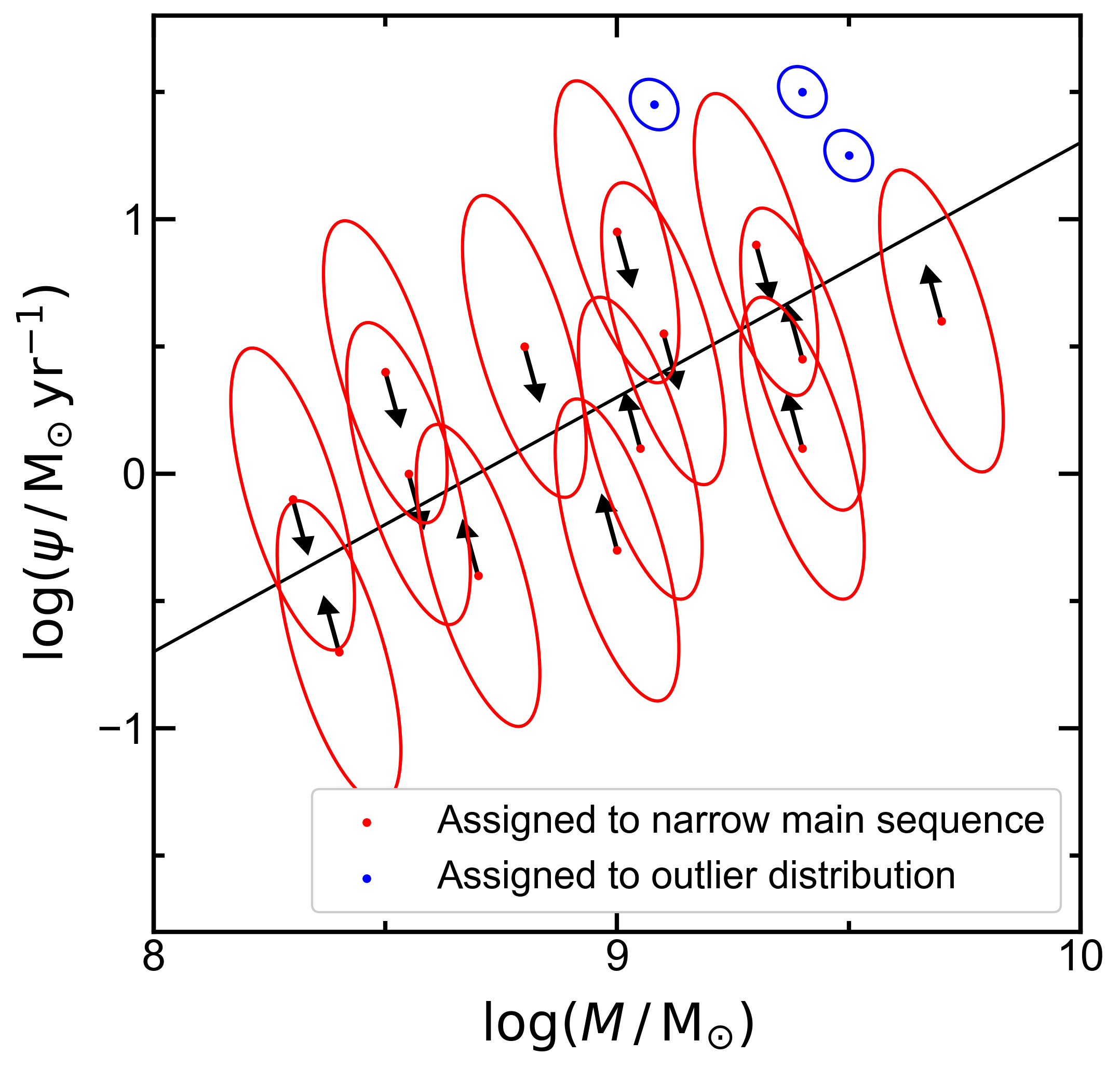}
    \caption{A simplified model of the main sequence (relation shown as a solid black line) showing $\log(\sfrInLog/\Msun\,\yr^{-1})$ plotted against $\log(\MstarInLog/\Msun)$. This cartoon displays the behaviour of the model when some objects on the main sequence have poor constraints that overlap (red points) while some have good constraints (blue points).  Even though all objects belong to the main sequence, in this case, with our model, the red points would be assigned to a main sequence with small intrinsic scatter while the objects with better constraints (blue points) would be assigned to the outlier model. In this scenario, the large overlap between the red objects will lead to shrinkage in the derived intrinsic scatter, as demonstrated by the black arrows.}
    \label{fig:low_scatter}
\end{figure}

The results for the intrinsic scatter in the three lowest redshift bins (Fig.~\ref{fig:redshift_bins}, bottom left panel) show that \OLnone\ measures the largest scatter, while the \OLhogg\ method shows the lowest measurements with a trend of decreasing scatter with increasing redshift. With further investigation of the $3<z<4$ redshift bin (that with lowest \OLhogg\ scatter measurement), we find that many of the objects are drawn to a tight main sequence relation with the \OLhogg\ method. This is because the co-varying uncertainties in \Mstar\ and \sfr\ (for the objects with high probability of being on the relation) approach the expected magnitude of the intrinsic scatter within the underlying relation. K07 demonstrate that in this regime, the method will under-estimate the intrinsic scatter. This demonstrates a potentially problematic feature of the \OLhogg\ model. In Fig.~\ref{fig:low_scatter} we show a simplified example where essentially, objects can be identified as belonging to the main sequence (red points) if they have posterior distributions that overlap in \MstarSFR\ space (effectively if the uncertainties are broad, as shown by red ellipses) while objects without overlapping posteriors (blue points) can be assigned to the outlier model. This will lead to shrinkage in the scatter about the derived main sequence (black arrows) by the amount allowed by the overlap. This behaviour may be particularly problematic if constraints on \Mstar\ and \sfr\ are poor, but led to occupy a similar region in \MstarSFR\ space by informative priors at the SED-fitting stage. We discuss this case further in Section~\ref{ss:Choice of Star Formation History}. Understanding this behaviour allows us to mitigate its effect when deriving the full redshift evolution model with the \OLhogg\ method (Section~\ref{ss:Redshift Evolution of the MS}).

The right three panels of Fig.~\ref{fig:redshift_bins} show the measured posterior medians and 68\% credible intervals for the parameters describing the fitted outlier distribution in the \OLhogg{} method: \OLprob{}, \OLmean{} and \OLscatter{}. The increase in \OLprob\ with redshift mirrors the decrease in \scatter\ with redshift for the lowest three redshift bins.  This demonstrates a degeneracy between these two parameters, and the importance of clearly identifying outlier galaxies for constraining \scatter\ of the main sequence. For clarity we have not plotted the redshift bins $4<z<5$ and $5<z<6$ in the panels displaying \OLmean\ and \OLscatter, as the probability of an object belonging to the outlier distribution, \OLprob, approaches zero in the highest redshift bins, leaving \OLmean\ and \OLscatter\ unconstrained. This, in turn, leads to all three methods measuring a similar main sequence slope ($\sim0.9, \sim1.3$), intercept ($\sim1.6, \sim2.4$) and intrinsic scatter ($\sim0.3, \sim0.4$) in the $4<z<5$ and $5<z<6$ bins respectively. We note that this may be due to the small sample sizes making outlier identification less secure, rather than there being fewer outliers in the underlying sample.

We have shown that the derived normalization and intrinsic scatter of the main sequence are highly sensitive to the presence and treatment of outliers in the data. In Section \ref{ss:Redshift Evolution of the MS} where we model the full redshift evolution of the main sequence, we choose to model the outlier population with the \OLhogg\ method as it is the \textit{only} method that can propagate the uncertainties on the treatment of outliers onto the parameters of interest. We note that the \OLscatter{} posterior probability is close to the prior lower-limit. The reason we impose a lower limit of $\OLscatter>1$ (as well as the upper limit in \OLprob) is to ensure that the outlier distribution does not account for objects primarily on the main sequence. As we are fitting two Gaussians to the population, a narrow outlier distribution introduces degeneracies. A wide outlier distribution also ensures that it is accounting for objects far from the main sequence. We will proceed with the current model as there are so few objects within the outlier population that adding further free parameters is unlikely to considerably improve the main sequence constraints. 

The results in this section also demonstrate that we cannot simultaneously constrain the intrinsic scatter and the outlier model within each redshift bin (especially within the $2<z<3$ and $3<z<4$ redshift bins), and so we account for this when constructing our full redshift-dependent model, as described in the following section.

\subsection{Redshift evolution of the MS}
\label{ss:Redshift Evolution of the MS}

Our Bayesian hierarchical model of the full redshift-dependent main sequence is composed of three levels. The first describes the distribution of stellar masses and redshifts:
\begin{equation}
    \begin{aligned}
        \Mstar \mid \Mdist &\sim \conditional{\Mstar}{\Mdist}\\
        \redshift \mid \zdist &\sim \conditional{\redshift}{\zdist}
        \label{eqn:level1}
    \end{aligned}
\end{equation}
\noindent
where we model $\conditional{\Mstar}{\Mdist}$ as a weighted linear combination of three Gaussians, called a Gaussian Mixture Model (GMM). The corresponding set of three means, standard deviations and relative weightings are denoted as \Mdist{}. We note that the mass distribution is not modelled as a function of redshift.  This does not mean that the \Mstar\ distribution within each redshift bin must look identical, but simply that the model learns the collapsed mass distribution of all objects regardless of their redshifts. We model $\conditional{\redshift}{\zdist}$ as a uniform distribution $\mathcal{U}(1.25,6)$ between our redshift limits. One would ideally also model the redshift distribution as non-uniform, potentially with another GMM. Modelling the redshift distribution explicitly would potentially help for deriving the relative likelihoods of peaks in probability that are separated significantly in redshift. However, in implementing this model we found that we had to handle these objects separately (as described later in this section). We therefore chose to not add more free parameters to the model at this level. 

Given \Mstar{} and \redshift{}, the second level of the model describes the probability distribution of \sfr\ as a linear combination of a main sequence distribution and an outlier distribution:
\begin{equation}
    \begin{aligned}
        \sfr|\Mstar,&\redshift \sim \PMS + \POL\\
        \PMS &= (1-\OLprob(\redshift))\,[\intercept(\redshift) + \slope(\redshift) (\Mstar-9.7) + \mathcal{N}(0, \scatter(\redshift)^2)]\\
        \POL &= \OLprob(\redshift)\,\,\mathcal{N}(\OLmean(\redshift),\OLscatter(\redshift)^2)\\
    \end{aligned}
\end{equation}
\noindent
where \intercept{}, \slope{}, \scatter{}, \OLprob{}, \OLmean{} and \OLscatter{} are now functions of redshift.

We determined suitable parametrizations for slope, intercept and scatter based upon the \OLhogg\ redshift bin measurements shown in Fig.~\ref{fig:redshift_bins}. The measurements of slope are consistent with being constant within the 68\% credible regions so we therefore choose to model it as constant:
\begin{equation}
    \slope(z) = \slope
    \label{eqn:slope(redshift)}
\end{equation}

The main sequence normalization, \intercept, is shown to increase with redshift from $z\sim1.25$ to $z\sim6$. We have shown that the normalization can be very strongly dependent on the modelling of the outliers and our redshift bin results are likely affected by this.  We see strong evolution in \intercept\ to higher redshifts that is not well reproduced by the parametrization of \cite{speagle_highly_2014}, and we do not trust the relatively low normalization of the $3<z<4$ bin compared to the $4<z<5$ bin for reasons described in Section~\ref{sss:redshift_bin_results}. We therefore chose to proceed with the physically motivated parametrization that follows the redshift evolution of the rate of accretion of gas onto dark matter halos \citep{birnboim_bursting_2007}.  We discuss the implications of this choice in Section~\ref{s:Discussion}:
\begin{equation}
    \intercept(z) = \log(\ssfrNorm (1+\redshift)^{\ssfrPower}) + 0.7\\
    \label{eqn:intercept(redshift}
\end{equation}
\noindent
where \ssfrNorm{} and \ssfrPower{} are the free parameters of our redshift-dependent model. The 0.7 is added for simplicity when plotting sSFR (at $\Mstar=9.7$) against redshift.

As discussed in Section \ref{sss:redshift_bin_results}, the measured value of the intrinsic scatter is strongly dependent on the treatment of outliers. We have demonstrated that, at least for redshift bins $2<z<3$ and $3<z<4$, the intrinsic scatter and the outlier model cannot be constrained independently. We therefore choose to parametrize the intrinsic scatter as a constant across all redshifts:
\begin{equation}
    \scatter(\redshift) = \scatter
    \label{eqn:scatter(redshift)}
\end{equation}
\noindent
This parametrization effectively allows the better \Mstar\ and \sfr\ constraints in the lowest redshift bin to constrain \scatter\ across all redshift bins. This choice is due to the limitations of the current data, which we discuss further in Section~\ref{ss:Choice of Star Formation History}.

\begin{figure*}
    \centering
    \subfigure{\includegraphics[width=0.45\textwidth]{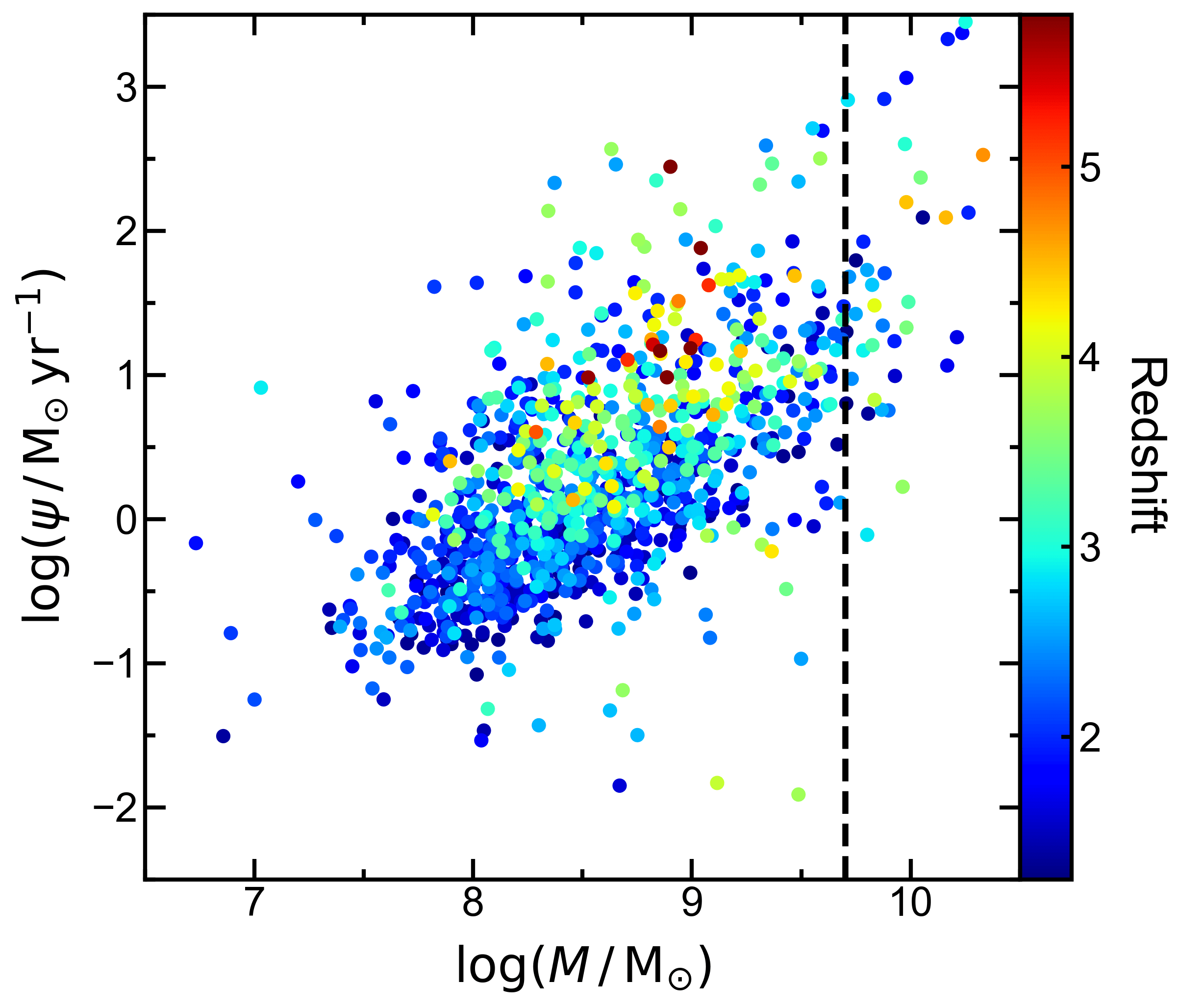}}
    \subfigure{\includegraphics[width=0.45\textwidth]{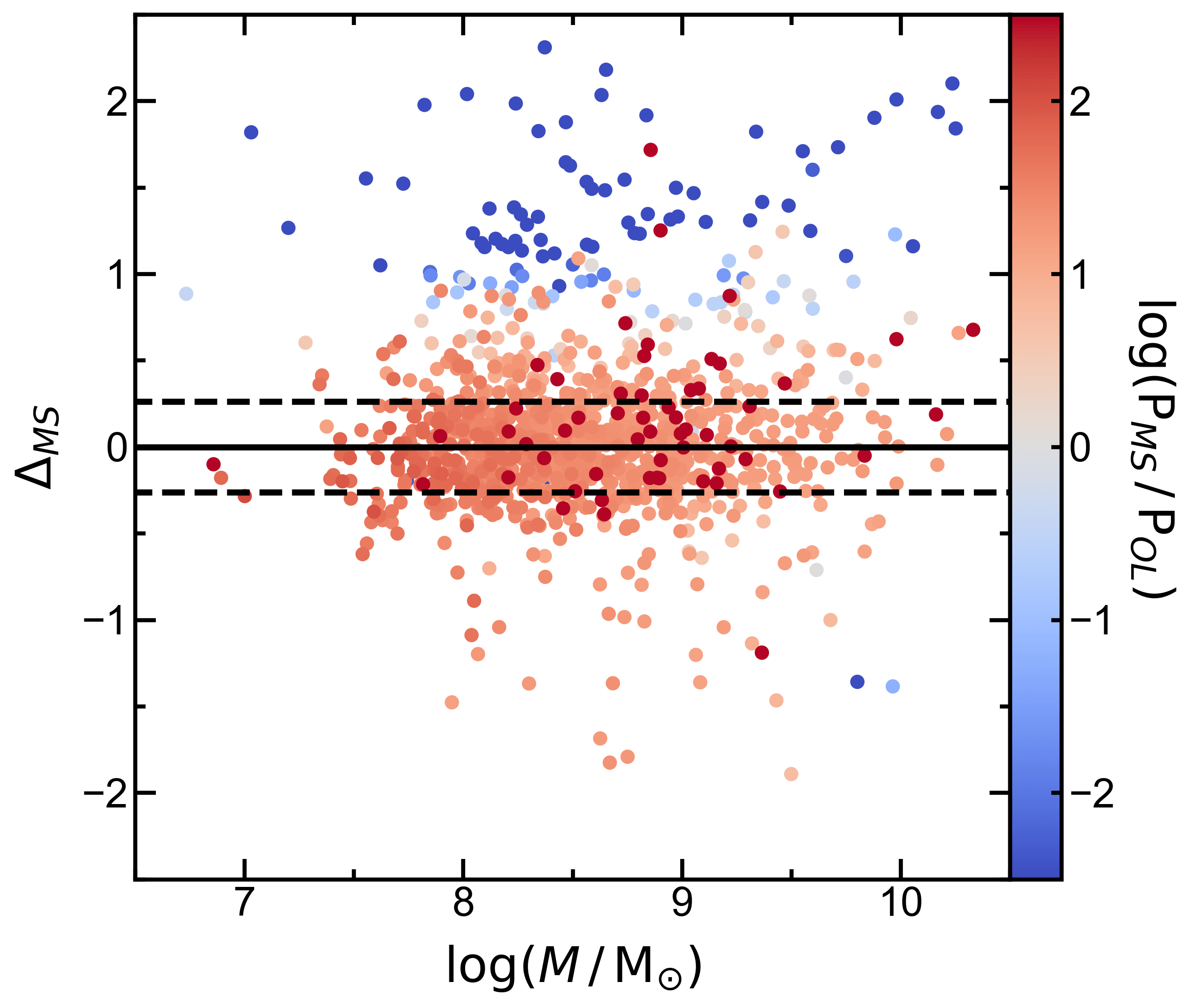}}
        \caption{\textit{Left panel:} \beagle{}-derived posterior median values of $\log(\MstarInLog/\Msun)$ and $\log(\sfrInLog/\Msun\,\yr^{-1})$ colour-coded by posterior median redshift for the objects in our sample. The vertical dashed line at $\log(\MstarInLog/\Msun)=9.7$ indicates the fixed mass at which the normalization in the main sequence is fitted. \textit{Right panel:} The same objects as in the left panel, showing the residual between $\log(\sfrInLog/\Msun\,\yr^{-1})$ and the fitted redshift-evolving main sequence, plotted against $\log(\MstarInLog/\Msun)$. The colour-coding shows the relative probability that the objects are on or off the main sequence.}
    \label{fig:MS_plots}
\end{figure*}

Fig.~\ref{fig:redshift_bins} (top right panel) shows a lack of obvious outliers at redshifts $z>4$. As guided by our \OLhogg{} method results, we opt to fit with constant \OLprob{}, \OLmean{} and \OLscatter{} for $1.25<z<4$, while at $z>4$ we make the assumption that all of our remaining objects are galaxies belonging to the star-forming main sequence. We implement this by setting \OLprob\ to zero at the highest redshifts:
\begin{equation}    
    \begin{aligned}
        \OLmean(\redshift) &= \OLmean \\
        \OLscatter(\redshift) &= \OLscatter \\
        \OLprob(\redshift) &=
            \begin{cases}
                \OLprob \, &\text{for $1.25<z<4$}\\
                0 \, &\text{for $z>4$}\\
            \end{cases}   
    \end{aligned}
\end{equation}
\noindent
It is worth noting here that our implementation of the outlier model is not the same as sigma-clipping. Each object has a probability of belonging to either the main sequence or the outlier distribution, in contrast to sigma-clipping which permanently removes outliers from the subset. However, an outlier model of this sort is not without potential risks. For example, at a given mass, as the star-forming main sequence evolves with redshift it will cross directly through the fixed outlier model. If the main sequence and outlier distributions were of comparable probability and width, the outlier distribution could have a similar (but not identical) effect to sigma-clipping of the relation (reducing the derived intrinsic scatter). We have attempted to mitigate this risk by constraining the relative probability of the outlier distribution to $\OLprob<0.5$ and its width to $\OLscatter{}>1$. Our complete model is then better described as a high probability main sequence, with a low probability distribution of outliers. To ensure that our derived main sequence is not dependent on our implementation of the outlier model, we additionally investigate the effect of using a uniform outlier distribution. The results of this test suggest that our measurement of the star-forming main sequence is robust, as further discussed in Section~\ref{s:Results}.

Originally, the third level of the K07 model accounts for the data and associated uncertainties, which are assumed to be point-wise estimates. Our data is one step further removed, being instead fluxes and flux uncertainties, rather than direct measurements of \Mstar, \sfr\ and \redshift. We follow the approach of CL21 (section 3.4) for dealing with this extra level of complexity. For simplicity of implementation, the individual object joint posteriors on \Mstar, \sfr\ and \redshift\ are modelled as a linear combination of three tri-variate Gaussians.

We found when implementing this model that the Gibbs sampler does not efficiently sample between peaks in posterior probability for a given object that are very far apart. It is beyond the scope of this work to re-visit the sampling method. Instead we use the information provided by the full sample to determine the most likely redshift peak in objects with peaks separated by $\Delta\redshift\geq2$. Each separate probability peak (determined from the Gaussians fitted to the \MstarSFRredshift\ posterior probability space) is multiplied by the probability that the object lies on the main sequence as measured within redshift bins (Fig.~\ref{fig:redshift_bins}), and integrated. The peak with higher probability overall is kept. Where two of the three Gaussians overlap within 1.5$\sigma$ in redshift, their probabilities are summed together and joint probability compared to the final peak. Where all three Gaussians overlap, no peak is rejected. Although only $\sim6$\% of galaxies had peaks removed, this was necessary because a small fraction of low-redshift objects with a peak at high redshift can significantly affect the results since there are far fewer high-redshift galaxies.

The implementation of the full model is based initially on the J. Meyers python implementation\footnote{https:\slash\slash github.com\slash jmeyers314\slash linmix} of the \kelly{} Gibbs sampler updated by \CL{} to accept GMM fits to posterior \MstarSFRredshift\ distributions derived from \beagle{} fitting. We release the code and input values used for this work\footnote{https:\slash\slash github.com\slash ls861\slash M-SFR-Sandles2022}.

\section{Results}
\label{s:Results}

In summary, our final model includes the following free parameters that we wish to constrain: the redshift evolution of the normalization at $\log(\MstarInLog/\Msun)=9.7$, parametrized by \ssfrNorm\ and \ssfrPower; the (redshift-independent) slope, \slope; the (redshift-independent) scatter, \scatter; and the parameters describing the outlier distribution, \OLprob{}, \OLmean{} and \OLscatter{}. Table~\ref{tab:kelly_priors} gives the parameters and priors used in this work (including our results as described in the following section).

\begin{table}
    \centering
    \caption{Parameters, fitted results (with 68\% credible intervals) and associated priors for the redshift-dependent main sequence model including the outlier distribution.}
    \label{tab:kelly_priors}
    \begin{tabular}{c c c}
        \toprule
        Parameter & $1.25<z<6$ & Prior \\
        \midrule
        \ssfrNorm & $0.12^{+0.04}_{-0.03}$ & $\mathrm{Uniform}\log \ssfrNorm \in [-3.0, 2.3]$ \\
        \ssfrPower & $2.40^{+0.18}_{-0.18}$ & $\mathrm{Uniform} \in [0.0, 5.0]$ \\
        \slope & $0.79^{+0.03}_{-0.04}$ & $\mathrm{Uniform} \in [-5.0, 5.0]$ \\
        \scatter & $0.26^{+0.02}_{-0.02}$ & $\mathrm{Uniform} \in [0.05, 5.0]$ \\
        $\OLprob^a$ & $0.19^{+0.03}_{-0.03}$ & $\mathrm{Uniform} \in [0.0, 0.5]$ \\
        \OLmean & $0.98^{+0.11}_{-0.10}$ & $\mathrm{Uniform} \in [-10.0, 10.0]$ \\
        \OLscatter & $1.01^{+0.02}_{-0.01}$ & $\mathrm{Uniform} \in [1.0, 10.0]$ \\
        \bottomrule
    \end{tabular}
    \vspace{1mm} \\ 
    $^a$Fixed to 0 for $z>4$. \hspace{50mm}
\end{table}

To constrain main sequence parameters (\ssfrNorm{}, \ssfrPower{}, \slope{}, \scatter{}, \OLprob{}, \OLmean{} and \OLscatter{}) for our subset of \totsample{} objects, we ran the full Bayesian hierarchical model for 20,000 iterations, four separate times. We checked for convergence between and within chains following the method described in chapter 11.4 of \cite{gelman_bayesian_2013}, ensuring an $\hat{R}$ value of $<1.1$. We use the second half of each chain (rejecting any burn-in phase) and combine them to determine the constraints on the parameters of interest.  The results are given in Table~\ref{tab:kelly_priors}.

\begin{figure}
    \centering
    \includegraphics[width=0.45\textwidth]{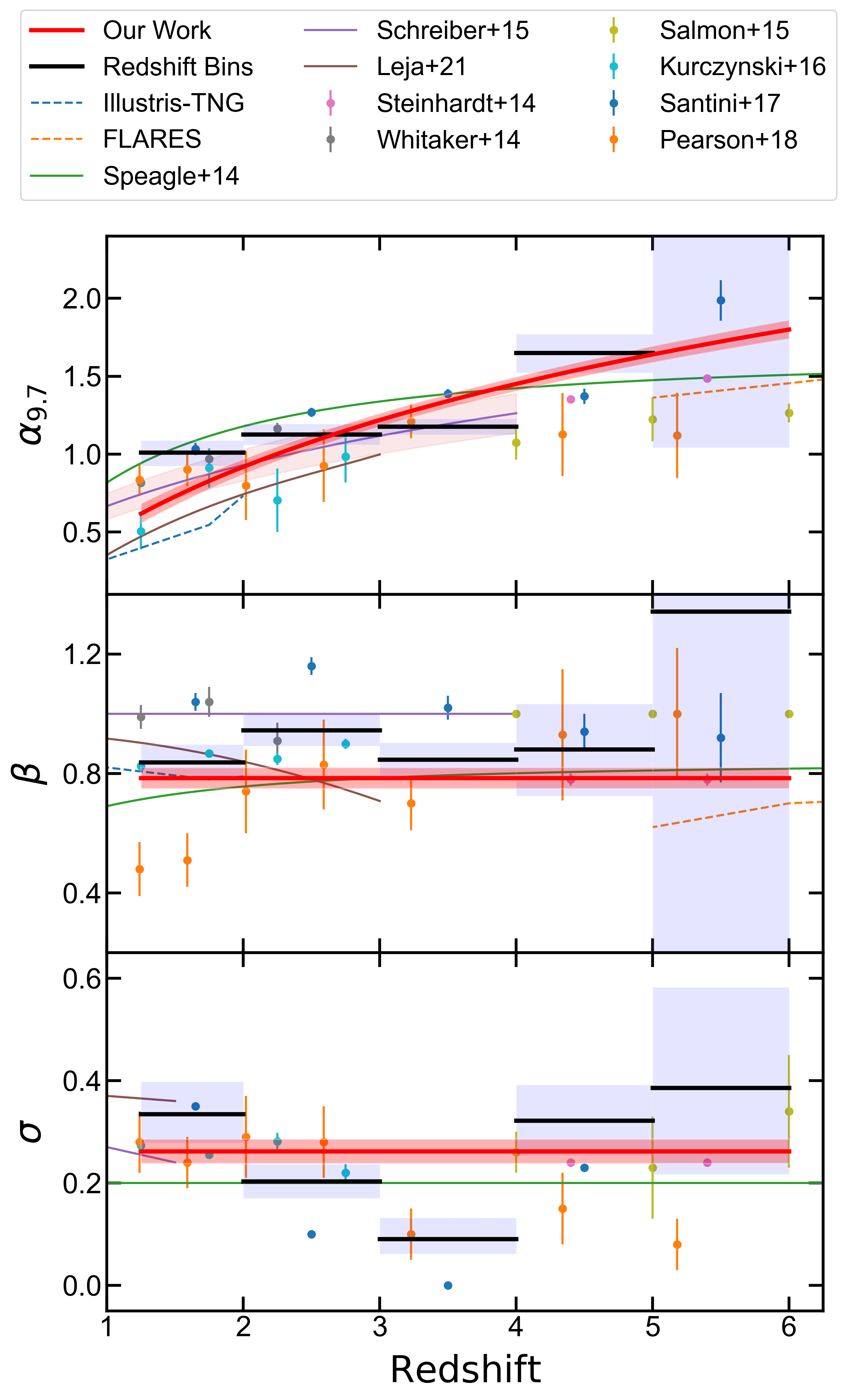}
    \caption{Redshift evolution of the normalization, \intercept{} (upper panel), slope, \slope{} (middle panel) and scatter, \scatter{} (bottom panel) of the main sequence. Red lines show the relations derived from the posterior medians of our fitted parameters. The red shaded regions show the 68\% credible intervals. For the case of \intercept{}, we sample from the joint posterior distribution of \ssfrNorm{} and \ssfrPower{}, before calculating the distribution of \intercept{} at each given redshift. The redshift bin results from the \OLhogg\ method (Fig. \ref{fig:redshift_bins}) are shown as solid black lines with 68\% credible intervals shaded blue. Results from the literature are also over-plotted following the legend.  Illustris-TNG results come from \protect\cite{donnari_erratum_2019} and FLARES results are taken from \protect\cite{lovell_first_2021} (values obtained via private communication). 
    All data in the top two panels is plotted for $\Mstar=9.7$.    Uncertainties on literature \intercept\ values are calculated in quadrature if the original work measured the intercept at a different mass. Where the fitted main sequence allows for curvature at high masses we plot the low mass linear slope \protect\citep[e.g.][]{schreiber_herschel_2015}.  For \protect\cite{leja_new_2021} we use broken power law parametrization fitted to the ridge in density in \MstarSFR\ space (their table 1).  For \protect\cite{whitaker_constraining_2014} we use the broken power law fit results. The scatter is sometimes measured as a function of mass. For \protect\cite{schreiber_herschel_2015}, \protect\cite{santini_star_2017} and \protect\cite{leja_new_2021} we plot the values of scatter taken at masses $\Mstar=10.2$, $\Mstar=9.2$ and $\Mstar=9.7$ respectively. We have converted the values where necessary to be consistent with a Chabrier IMF.}
    \label{fig:fullrun_vs_redshift}
\end{figure}

We display the results in Fig.~\ref{fig:MS_plots}. 
The left panel shows the \beagle{}-derived posterior median \Mstar{} and \sfr{} plotted in the \MstarSFR\ plane colour-coded by redshift. We see a clear sign of an increase in the normalization with redshift, consistent with previous literature results. The dashed black line passing through $\Mstar=9.7$ indicates the mass at which we define the normalization of the main sequence, \intercept{}. The right panel shows the offsets from the fitted redshift-evolving main sequence vs. \Mstar, colour-coded by the logarithm of the ratio of the probability that objects are on or off the main sequence. A value of zero corresponds to an equal likelihood that the object is on or off the main sequence. In our model, objects with a posterior median redshift value of $z>4$ are assigned a zero probability of belonging to the outlier distribution and are shown as dark red circles with artificially assigned values of $\log(\PMS/\POL)=3$.

\begin{figure*}
    \centering
    \includegraphics[width=0.7\textwidth]{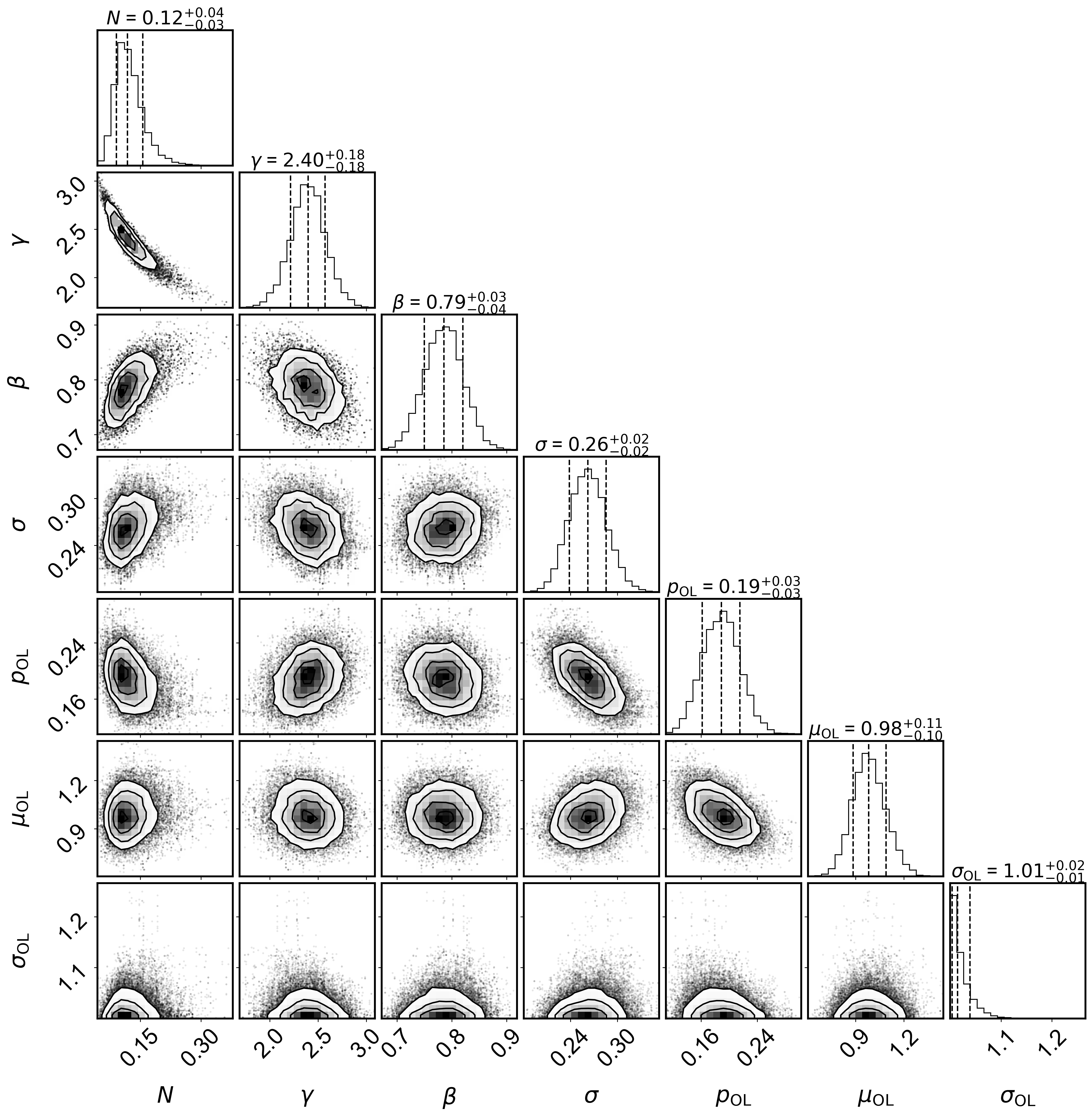}
    \caption{\textit{Main diagonal:} Marginal probability distributions for each of the seven fitted parameters (\ssfrNorm{}, \ssfrPower{}, \slope{}, \scatter{}, \OLprob{}, \OLmean{} and \OLscatter{}) in our full redshift dependent run. Vertical dashed lines show the median and 68\% credible intervals. \textit{Off-centre:} Joint posterior distributions for every pair of parameters with solid black lines to show the 1, 2 and $3\sigma$ contours.}
    \label{fig:corner_full_run}
    \centering
    \includegraphics[width=0.9\textwidth]{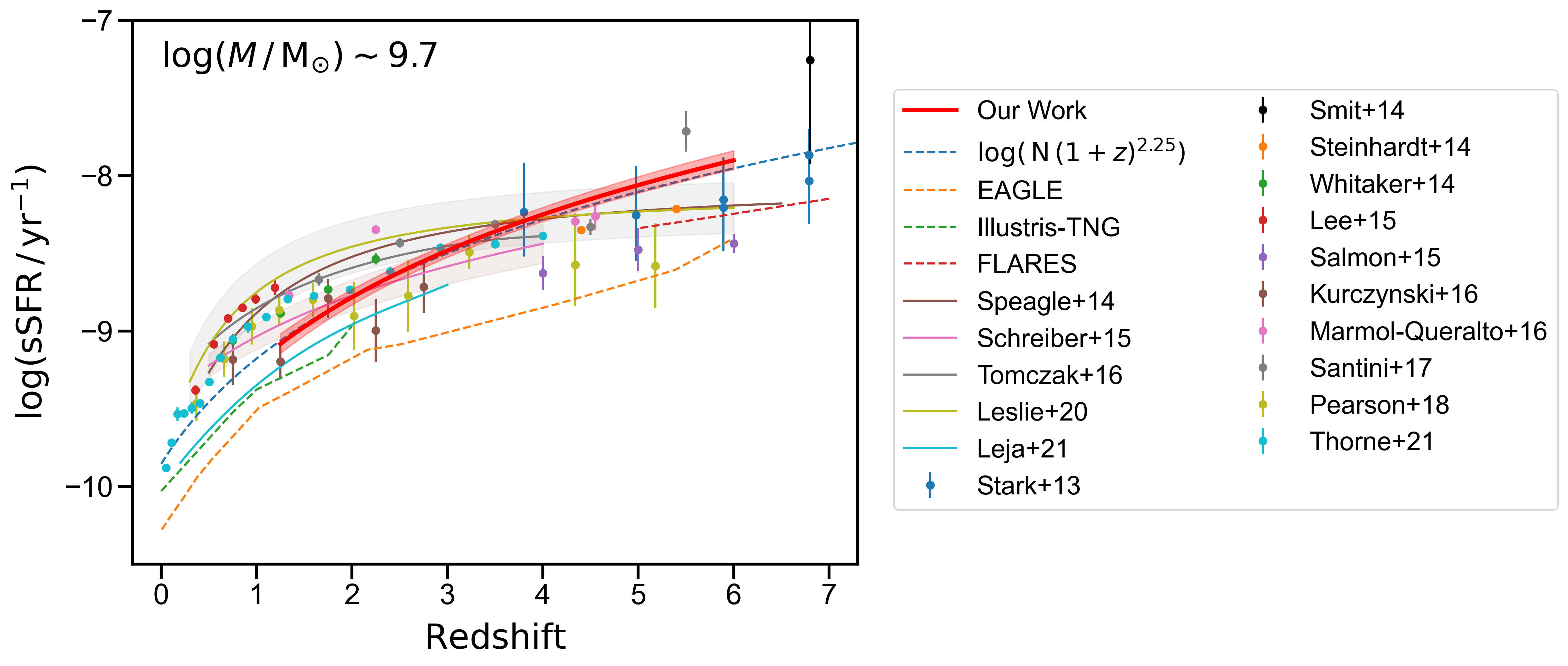}
    \caption{The evolution of log(sSFR) plotted as a function of redshift for galaxies with a stellar mass of $\sim 10^{9.7}\,\Msun$. Our work is shown by the solid red line with shaded 68\% credible intervals (as described in caption of Fig.~\ref{fig:fullrun_vs_redshift}. Coloured circles show previous observational results from the literature. The results for \protect\cite{leslie_vla-cosmos_2020} and \protect\cite{tomczak_sfr-m_2016} are taken from the relations fitted only to star-forming galaxies. Results from the EAGLE simulations (Ref-L100N1504 model) are taken from \protect\cite{furlong_evolution_2015}, Illustris-TNG from \protect\cite{donnari_erratum_2019}, and FLARES from \protect\cite{lovell_first_2021}. The dashed blue line represents a simple functional form consistent with the evolution of the accretion rate of gas onto parent halos, normalized to our work at $z=2$.}
    \label{fig:ssfr_vs_redshift}
\end{figure*}

The objects with poorly constrained \sfr\ that are rejected prior to fitting are not plotted, leaving more objects significantly above the relation than below. The outlier model primarily accounts for objects above the main sequence but is broad enough to also encompass those below the relation.

Fig.~\ref{fig:fullrun_vs_redshift} shows the derived posterior median values and 68\% credible intervals (red line and shaded regions, respectively) of the main sequence slope, redshift-dependent normalization and intrinsic scatter spanning $1.25<z<6$. The \OLhogg\ redshift bin values are shown as solid black lines and shaded blue regions. The panels additionally include data obtained from the literature\footnote{We ensure consistency of IMF using conversion factors $-0.21$ for Salpeter to Chabrier \Mstar{}, $-0.20$ for Salpeter to Chabrier \sfr{} and $-0.03$ for Kroupa to Chabrier \Mstar{} and \sfr{}.} and simulations. The top panel shows the redshift evolution of \intercept. The $1.25<z<2$ redshift bin results are higher than those of the full model. This is a known problem with our full model parametrization. It is based on the measured accretion rate of gas onto parent halos, but at low redshift many studies measure higher SFRs than accounted for by this evolution.  The recent work of \cite{leja_new_2021} find a much lower normalization for the main sequence at low-to-intermediate redshifts, agreeing well with the predictions from hydrodynamic simulations of galaxy formation \citep[e.g. Illustris-TNG][]{donnari_star_2019}.  Our measured normalization is still higher than that measured by \cite{leja_new_2021}, which can be attributed to the different SFHs employed (their non-parametric histories show older, more massive galaxies than estimates derived with simple analytic forms like the DE used here).  We discuss further the limitations of our results with respect to SFH in Section~\ref{ss:Choice of Star Formation History}.  The high-redshift bins are driving the fit of the redshift evolution of \intercept, showing that our results are inconsistent with a flatter evolution at high redshifts as measured by e.g. \cite{speagle_highly_2014} and \cite{pearson_main_2018}.

The middle panel of Fig.~\ref{fig:fullrun_vs_redshift} shows the measurements of main sequence slope, \slope{}, across the full redshift range. Our measurements agree well with those of \cite{kurczynski_evolution_2016}, \cite{speagle_highly_2014} and \cite{pearson_main_2018} above $z>2$, but are somewhat shallower than those measured by \cite{santini_star_2017} and \cite{leja_new_2021}. The \cite{schreiber_herschel_2015} and \cite{salmon_relation_2015} slope values are fixed (where we take the low mass slope of the curved relation fitted in \citealt{schreiber_herschel_2015}, and we have chosen to plot the results from \citealt{salmon_relation_2015} fitted with a fixed slope). We discuss in Section~\ref{ss:Choice of Star Formation History} the effects of the priors employed in \beagle, which will strongly affect the measured slope.

The bottom panel of Fig.~\ref{fig:fullrun_vs_redshift} shows measurements of the scatter about the main sequence.  Our constant, intrinsic scatter estimate, \scatter, agrees well with \cite{pearson_main_2018} up to $z\sim3$, \cite{kurczynski_evolution_2016}, \cite{steinhardt_star_2014} and \cite{salmon_relation_2015}. We note that the value of scatter plotted for \cite{steinhardt_star_2014} is an observed scatter, rather than the intrinsic value. Interestingly, some of the studies show decreasing scatter with particularly low estimates at $z\gtrsim3$ \citep[e.g.][]{pearson_main_2018, santini_star_2017} which agree better with our $z\sim3$ redshift bin results. However, at these redshifts the \cite{pearson_main_2018} main sequence has a very strong lower-limit that appears to be biasing the scatter to low values (see their figure 8).  Additionally, \cite{santini_star_2017} used the same data set as us, and so are likely inhibited by the same limitations in constraints on \Mstar\ and \sfr, which would under-estimate the \scatter\ at $z\sim3$.

As discussed in Section~\ref{sss:Modelling Outliers}, it was important to ensure that our derived main sequence was not strongly dependent on our implementation of the outlier model. As a simple check, we decided to also fit the full $1.25<z<6$ subset with an adjusted outlier model: a truncated Gaussian between $-2.0<\sfr<3.75$ with fixed $\OLmean=0.0$ and $\OLscatter=9.0$ (effectively a uniform outlier distribution between $-2.0<\sfr<3.75$ for $z<4$). The redshift evolution of the fitted main sequence intercept ($\ssfrNorm=0.16\pm^{0.05}_{0.04}$ and $\ssfrPower=2.29\pm^{0.19}_{0.19}$) remained consistent with the original model ($\ssfrNorm=0.12\pm^{0.04}_{0.03}$ and $\ssfrPower=2.40\pm^{0.18}_{0.18}$), whilst the slope was measured to be only slightly lower at $\slope=0.71\pm^{0.04}_{0.04}$. The intrinsic scatter was determined to be significantly higher: $\scatter=0.46\pm^{0.03}_{0.04}$. We note that this was primarily due to the uniform outlier model being assigned a low probability ($\OLprob=0.04\pm^{0.02}_{0.02}$), with most of the objects having a high probability of being on the main sequence. This suggests that a uniform outlier model is inadequate for describing the outlier population.

Fig.~\ref{fig:corner_full_run} shows the bi-variate posterior distributions between each pair of parameters. The main diagonal shows the marginalized posterior distributions of each individual parameter, with vertical lines representing the median and 68\% credible intervals.  This helps to understand where degeneracies between different parameters may impact our results.   We see a negative correlation between \scatter\ and \OLprob, \OLprob\ and \OLmean.  This further demonstrates the sensitivity of our \scatter\ estimates to the details of the outlier model.  We also see strong degeneracies between \ssfrPower{}, \ssfrNorm{} and \slope, such that low \ssfrNorm\ requires a low \slope, but high \ssfrNorm.  This might explain why we measure shallower slope in the full model compared to in the redshift bins.

Fig.~\ref{fig:ssfr_vs_redshift} shows the redshift evolution of the specific star formation rate (sSFR) at $\log(\MstarInLog/\Msun)=9.7$. By definition, at a fixed stellar mass, sSFR follows the redshift evolution of the main sequence normalization at that mass.  Measurements of sSFR are not always derived from measuring the main sequence, and so we can compare to more results in the literature. Our measured evolution is clearly more consistent with the data that show significantly higher sSFR at high redshifts, compared to the data that suggest a flatter evolution to high redshift. Our measurement of \ssfrPower\ ($2.40^{+0.18}_{-0.18}$) is consistent with the evolution of the accretion rate of gas onto parent halos (shown as the dashed blue line, with evolution $\sim(1+z)^{2.25}$).

\begin{figure}
    \centering
    \includegraphics[width=0.45\textwidth]{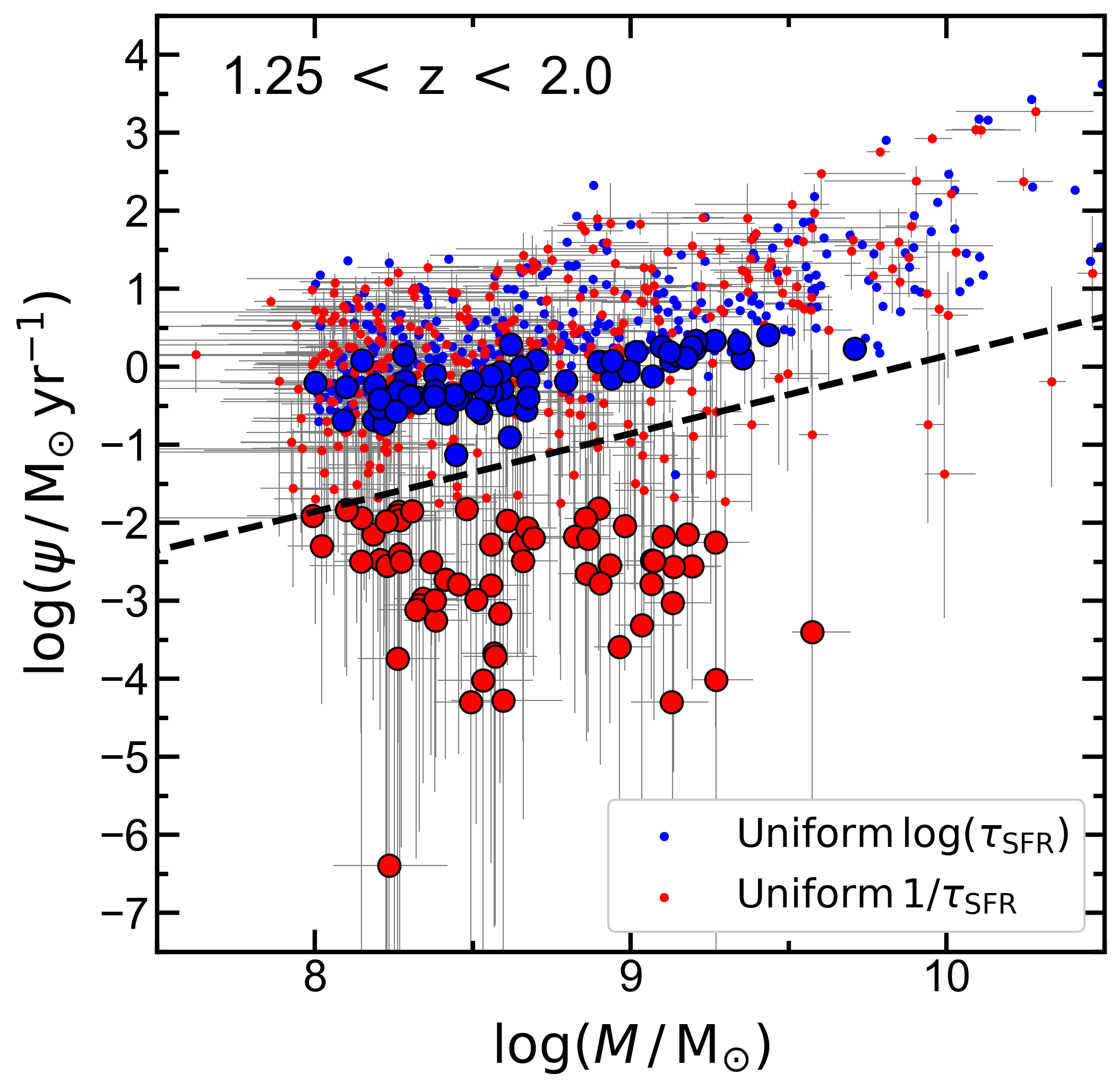}
    \caption{SFR vs. stellar mass for objects within the $1.25<z<2$ bin measured with different priors on \tausfr.  We display the SFR and mass constraints before correcting for magnification to more clearly show the effects of the priors.  The blue points show the measurements used in our fiducial model, fitted with a uniform prior on $\log(\tausfr)$, while the red points show the results when fitted with a uniform prior on $1/\tausfr$. Grey error bars showing 68\% credible intervals in $\log(\MstarInLog/\Msun)$ and $\log(\sfrInLog/\Msun\yr^{-1})$ are shown for the red points (for clarity we do not plot uncertainties for the original estimates). Large circles show the objects that, when fit with uniform prior on $1/\tausfr$, give $\log(\sfrInLog/\Msun\yr^{-1})<-1.8$.  The black dashed line shows the limit at which the prior density falls off quickly for the prior used in our fiducial model (see Fig.~\ref{fig:SFH_prior_heatplots}).}
    \label{fig:new_SFH_prior}
\end{figure}

\section{Discussion}
\label{s:Discussion}

\subsection{Choice of star formation history}
\label{ss:Choice of Star Formation History}

\subsubsection{Determining constraints on \Mstar\ and \sfr}
\label{sss:Determining constraints}

\begin{figure*}
    \centering
    \includegraphics[width=0.9\textwidth]{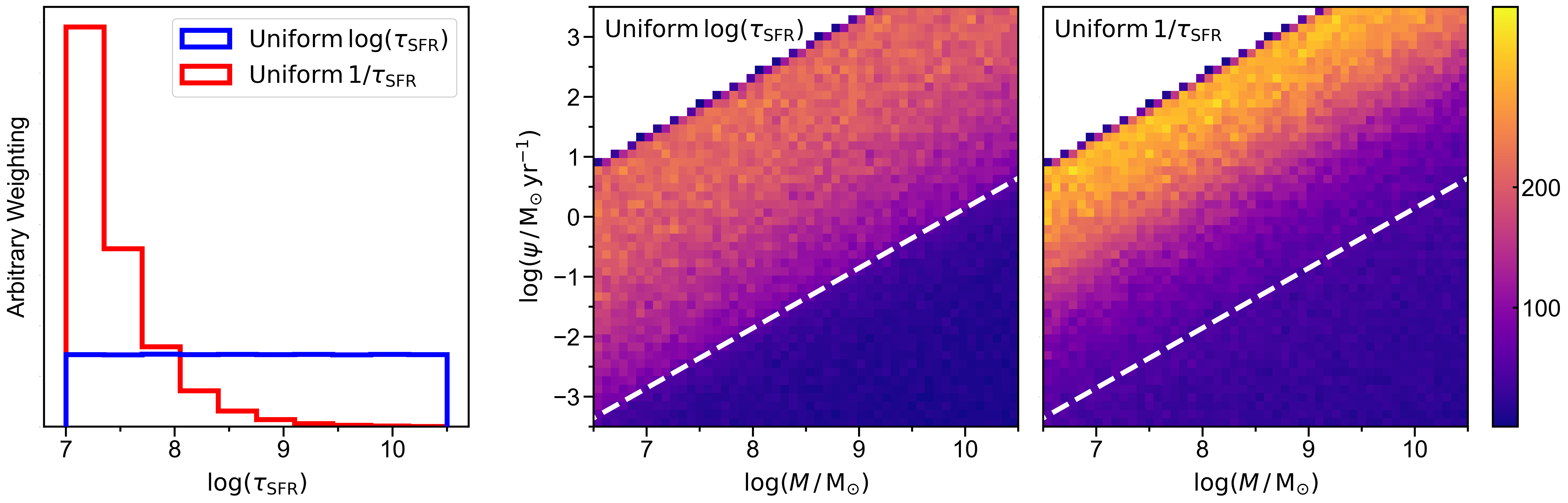}
    \caption{The two priors on \tausfr\ employed when testing the dependence of the results in the $1.25<z<2$ bin on the priors employed in the fits.  The fiducial prior employed for our fits is uniform on $\log(\tausfr)$, and we also fit with a prior that is uniform on $1/\tausfr$.  The two priors are plotted in the left panel. The middle panel shows the prior probability weighting in the \MstarSFR\ plane for our fiducial prior on \tausfr, while the uniform prior on $1/\tausfr$ is shown in the right panel.  The colour-coding shows the weighting in the prior with arbitrary normalization.  The white dashed line shows the lower limit in the prior space imposed by the rising portion of the DE SFH (see text for details).  As can be seen, the weighting in our fiducial prior (middle) falls significantly below this line.}
    \label{fig:SFH_prior_heatplots}
\end{figure*}

In Section \ref{s:Results} we have presented our measurement of the slope, intercept and scatter of the star-forming main sequence between redshifts $1.25<z<6$. We performed the \beagle\ fits with a delayed exponentially declining SFH of the form $\sfrInLog(\msa) \propto \msa \, \exp(-\msa/\tausfr)$. CL21 demonstrated that the constraints on \tausfr\ can be poor, and subsequently lead to significantly biased estimates on main sequence parameters. The CL21 study was based on simulated data at $z\sim5$, using a set of \JWST\ Near-infrared camera (NIRCam) filters. Our data-set spans a wide redshift range and uses a different set of filters, but we can still evaluate the possible impact of poorly constrained SFH parameters on the derived main sequence by comparing fits performed using different priors on \tausfr. We therefore re-ran \beagle\ using a uniform prior on $1/\tausfr$ (a prior suggested by \citealt{carnall_how_2019}), within the same limits as our fiducial prior, which was uniform on $\log(\tausfr)$ (see Table~\ref{tab:beagle_priors}). 

The results for the \MstarSFR\ plane for the $1.25<z<2$ redshift bin are shown in Fig.~\ref{fig:new_SFH_prior}, where we plot the values prior to correcting for any magnification correction to more clearly display any effects from the priors. The original posterior medians are shown as blue points, while the new posterior medians are coloured red, and their 68\% credible regions in \Mstar\ and \sfr\ are displayed as grey error bars (for clarity we do not plot uncertainties for the original estimates). We see a large excess of red points significantly below the relation with very large uncertainties in \sfr. Objects with $\log(\sfrInLog/\Msun\,\yr^{-1})<-1.8$ are shown as large circles, as are the corresponding objects fitted with the original prior. 
As a comparison, we applied the method described in Section~\ref{ss:Redshift Bins} to derive a main sequence slope, intercept and scatter for the red objects. As one may expect from visually inspecting Fig.~\ref{fig:new_SFH_prior}, we measured a steeper slope of $0.93^{+0.09}_{-0.09}$ (compared with $0.84^{+0.06}_{-0.06}$) and a larger intrinsic scatter of $0.67^{+0.04}_{-0.04}$ (compared with $0.33^{+0.06}_{-0.06}$). The newly fitted main sequence intercept ($1.15^{+0.10}_{-0.11}$) was consistent with that of the original run ($1.01^{+0.08}_{-0.09}$). However, the presence of an outlier distribution was significantly down-weighted with $\OLprob=0.00^{+0.01}_{-0.00}$ (compared with $0.12^{+0.06}_{-0.06}$).

To understand the behaviour of the fits with uniform prior on $\log(\tausfr)$, we need to visualize the resulting prior in \MstarSFR\ space. This is shown in Fig.~\ref{fig:SFH_prior_heatplots}. We see a ``ridge'' in the original prior, which is very close to the values of \sfr\ measured for objects that are subsequently measured to have very low \sfr\ with the new prior (we have plotted the posterior medians, so they are slightly above the ridge which would represent the extent of the 95\% credible intervals). The prior in \MstarSFR\ space is, in fact, a combination of how the priors on \tausfr\ and \msa\ interact. When $\msa<\tausfr$, the SFH is in the rising portion, prior to the exponential decline. This rising history actually has a hard lower limit in \MstarSFR\ shown by the dashed line in Fig~\ref{fig:new_SFH_prior} and Fig~\ref{fig:SFH_prior_heatplots} (middle and right panels). The uniform prior on $\log(\tausfr)$ has larger weight in high \tausfr\ values (left panel, Fig.~\ref{fig:SFH_prior_heatplots}), which puts higher weighting into the rising portion of the SFH. This can lead to uncertainties on \sfr\ that are small, suggesting that \sfr\ has been constrained by the data, when in fact the small uncertainties are caused by the informative prior on \tausfr. 

We used a redshift-dependent mass completeness cut (Fig.~\ref{fig:lower_mass_limit}) to determine which objects we would use to measure the main sequence. This simple exercise demonstrates that for low mass objects (still above the mass completeness cut), the signal-to-noise ratio is insufficient to constrain \sfr. To obtain results of the main sequence that are not dominated by the priors on SFH parameters, one needs to determine the mass limit at which \Mstar\ and \sfr\ are both constrained to a certain level of accuracy. Within this redshift bin, a by-eye assessment (from Fig.~\ref{fig:new_SFH_prior}) suggests that a lower mass limit of $\log(\MstarInLog/\Msun)\gtrsim9.3-9.5$ would be appropriate, more than an order of magnitude higher than our mass-completeness limit. 

An alternative approach that would allow deriving constraints on the main sequence to lower stellar mass, would involve censoring the data with poor constraints on \sfr{}. This might take the form of retaining only objects with \sfr{} above a lower limit which is defined by how well \sfr{} is constrained. This type of modelling is explicitly accounted for in the package \leopy, \citep{feldmann_leo-py_2019}, but the underlying model does not explicitly account for outliers. This is mitigated by \cite{feldmann_leo-py_2019} when fitting to the main sequence at $0.01<z<0.05$ by modelling the distribution as asymmetrically distributed about the main sequence, allowing a tail to low \sfr\ to account for quiescent galaxies and those in transition. However, we have demonstrated that objects above the relation also need to be accounted for in an outlier model. It is beyond the scope of this paper to include modelling of censored data, which we defer to future work.

It is possible that the lower-limit in the prior for a rising SFH is reducing the slope we measure from the main sequence, since the region of low \sfr\ at low mass required to produce a steeper slope is relatively inaccessible thanks to the effective prior. This is likely why our study and that of \cite{kurczynski_evolution_2016}, who use the same SFH and similar prior (priv. communication) on $\tausfr$, measure a lower slope than \cite{santini_star_2017}. \cite{santini_star_2017} measure \Mstar\ with a similar SFH to ours (but constrained to the rising portion at $z>4$), but estimate \sfr\ indirectly using assumptions of a constant \sfr\ (via the \citealt{kennicutt_star_2012} UV-luminosity to \sfr\ calibration). This breaking of the dependence of \sfr\ and \Mstar\ on the same SFH can reduce the impact, somewhat, of hard lower- and upper-limits in the \MstarSFR\ plane imposed by priors.  However, the individual \Mstar\ and \sfr\ estimates are still limited by their respective priors and assumptions. 

\begin{figure}
    \centering
    \includegraphics[width=0.45\textwidth]{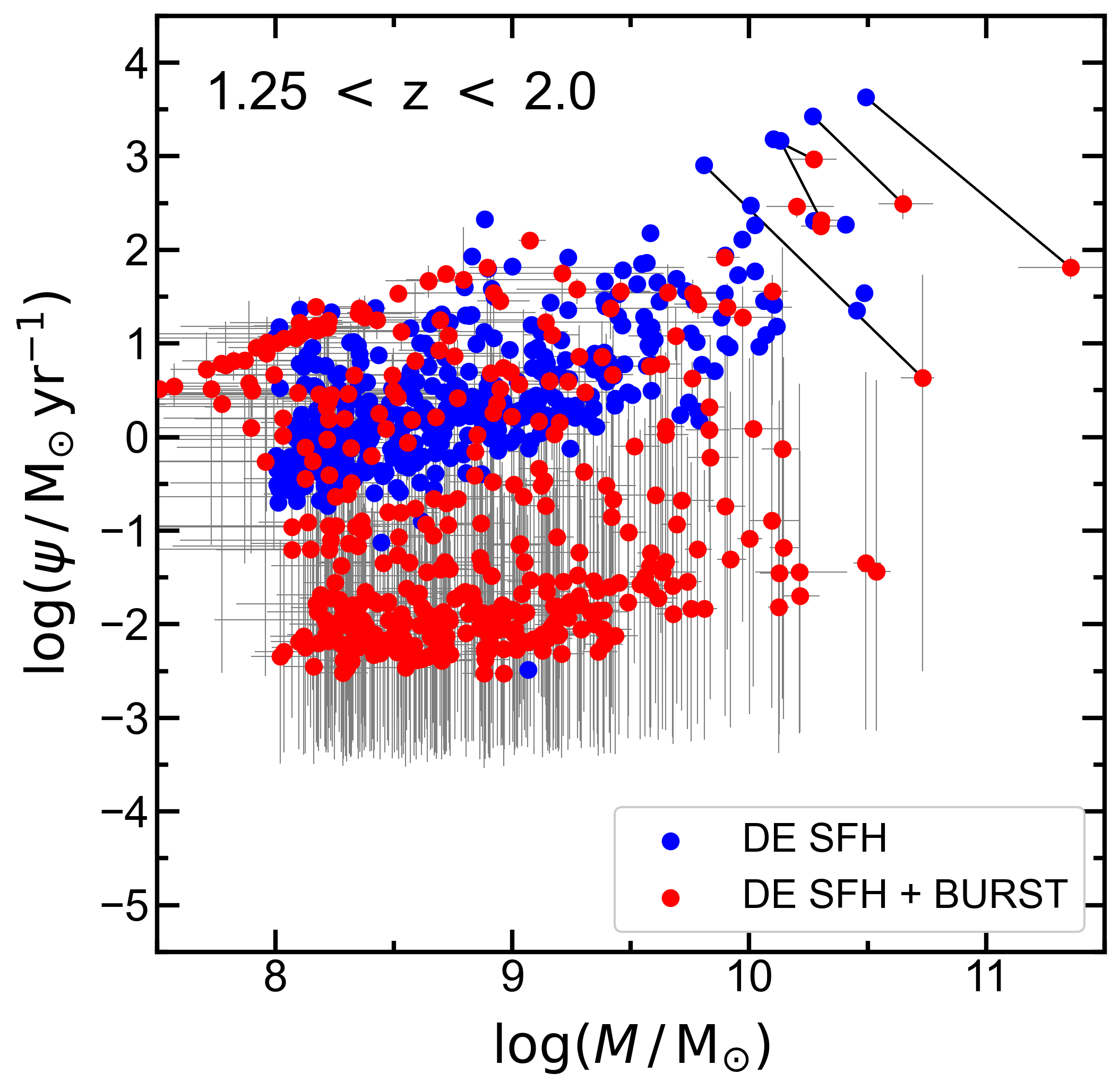}
    \caption{SFR vs. stellar mass for objects in the $1.25<z<2$ bin for objects fitted with two different SFHs. As for Fig.~\ref{fig:new_SFH_prior}, we display the SFR and mass constraints before correcting for magnification as it more clearly shows the effects of unconstrained parameters. The blue points show posterior medians measured when fitting with a delayed exponential (DE) SFH, while the red points show the posterior medians measured with a DE SFH where the SFR within the most recent $10\,\Myr$ is constant, and allowed to vary independently of the previous history (DE SFH + BURST).  For clarity we show the 68\% credibility regions for the DE + BURST measurements.  The black lines connect the measurements for 5 object originally identified as residing above the main sequence relation with the DE SFH, that sit on or below the main sequence when fitted with a DE + BURST SFH. }
    \label{fig:new_SFH}
\end{figure}    

\subsubsection{Form of SFH}

Our chosen SFH is still very constraining in form; it ties the current SFR directly to the past star formation. There is an argument, often used, that the rest-frame ultra-violet varies with SFR on a timescale of $\sim100\,\Myr$, and so broad-band photometry is not sensitive to short timescales of star formation. By this argument, short timescales do not have to be represented in the SFH when fitting only to broad-band photometry. This assumption is clearly incorrect at high redshift where emission lines (sensitive to \sfr\ on timescales $\sim10\,\Myr$) have been demonstrated to significantly affect broad-band fluxes \citep{curtis-lake_ages_2013, stark_keck_2013, de_barros_properties_2014, smit_evidence_2014, curtis-lake_modelling_2021}. New studies with more complex SFHs demonstrate how measurements made with simpler SFH prescriptions can be biased \citep{carnall_how_2019, leja_how_2019}. \cite{leja_older_2019} derive older ages and lower SFRs when fitting to multi-band photometry from the 3D-HST catalogues \citep{skelton_3d-hst_2014} with a SFH described by discrete bins of star formation. This leads to a lower measurement of the normalization of the main sequence \citep{leja_new_2021}.

\begin{figure}
    \centering
    \includegraphics[width=0.45\textwidth]{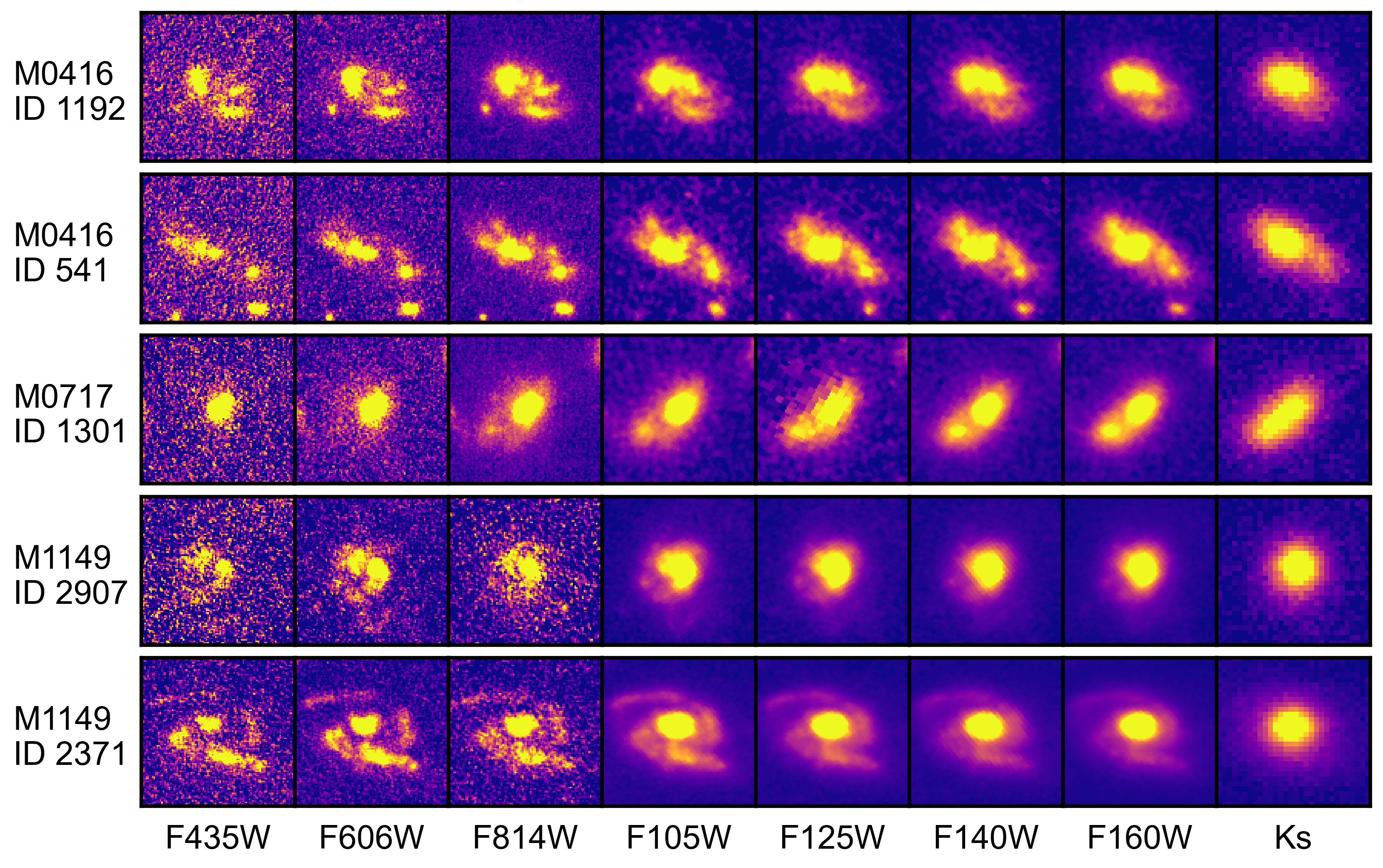}
    \caption{Postage stamps of the five objects originally identified as outliers at high \Mstar\ and \sfr\ in the $1.25<z<6$ bin (see Fig.~\ref{fig:new_SFH}). The objects increase in \beagle{}-derived \Mstar\ (prior to magnification correction) with the DE SFH from top to bottom.}
    \label{fig:panels.png}
\end{figure}    

We have shown that our SFR estimates for objects with $\log(\sfrInLog/\Msun\,\yr^{-1})\lesssim0$ in the $1.25<z<2$ bin are highly dependent on SFH priors. However, for objects with firm \Mstar\ and \sfr\ constraints with the DE SFH, it can be instructive to fit with a SFH with more freedom. We fit with a simple history that completely decouples the present SFR with the previous SFH (and hence with the accumulated stellar mass). This SFH describes the current star formation with a constant history over the last $10\,\Myr$ (with a uniform prior between $-4<\log(\sfrInLog/\Msun\,\yr^{-1})<4$) while earlier times are described by a DE. We label it DE SFH + BURST. The results are displayed in Fig.~\ref{fig:new_SFH} as red points, with the 68\% credible regions in \Mstar\ and \sfr\ shown as grey error bars.  We plot the constraints prior to correcting for magnification. The blue points show the original posterior median constraints. As expected, given more freedom in the SFH, a very large fraction of the objects have very poorly constrained \sfr\ (low SFR objects with large uncertainties in SFR), while some objects also have very poorly constrained masses (those with high \sfr\ that also have large uncertainties on \Mstar).
When further analysing the DE SFH + BURST sample in order to derive a main sequence slope, intercept and scatter, we determined that a cut of 1.0 dex in \sfr\ uncertainty would be necessary to remove the highly unconstrained objects. For the DE SFH + BURST sample in the $1.25<z<2$ redshift bin, this flagged 83\% of the objects. We therefore did not proceed to fit a main sequence to this subset. For comparison, only 13\% of the original DE SFH sample had uncertainties in \sfr\ greater than 1.0 dex.

\begin{figure*}
    \centering
    \includegraphics[width=0.9\textwidth]{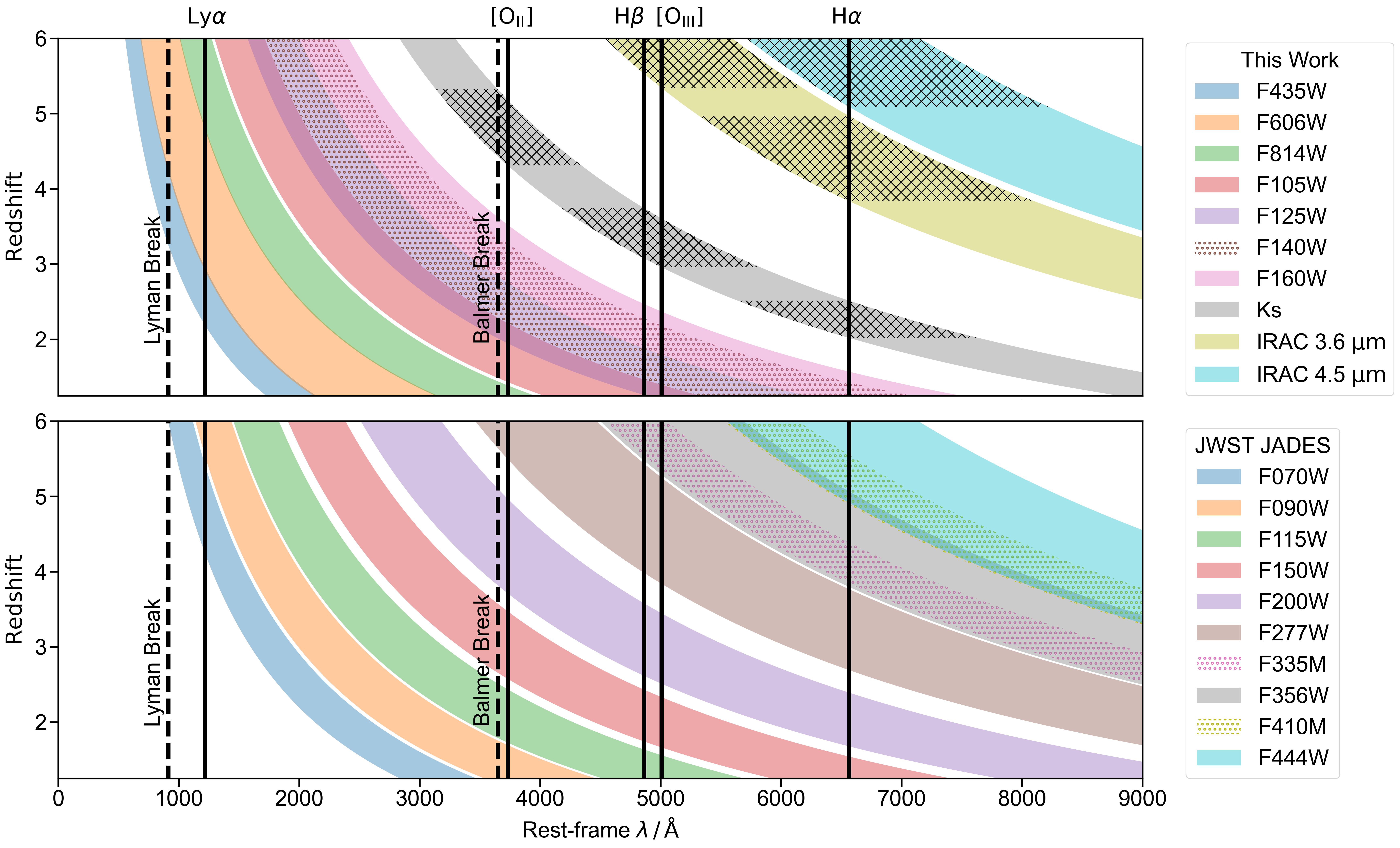}
    \caption{Redshift vs. rest-frame wavelength showing the filter coverage over key spectral features. The dashed black lines from left to right show the Lyman and Balmer breaks. The solid black lines from left to right represent \La, \oii, \Hb, \oiii\ and \Ha. 
    \textit{Top panel:} The shaded regions represent the \astrodeep\ filter-set. The dotted area shows the F140W filter which overlaps the F125W and F160W filters. The black hatched regions show the redshifts at which the Ks and IRAC filters are contaminated by strong emission lines. 
    \textit{Bottom panel:} The shaded regions represent the JADES broad-band filter-set. The dotted regions show the two JADES medium-band filters, F335M and F410M.}
    \label{fig:emission_lines}
\end{figure*}

For certain high SFR objects that are originally identified as outliers above the main sequence, the DE SFH could not allow for a recent burst of star formation without forcing the galaxy to be very young ($\msa\sim10^{7.2}$). We demonstrate the trajectory of five such outliers when fitting with the more flexible SFH by black lines connecting the original posterior medians (blue points) with the new estimates (red points). When the SFH allows the current SFR to be decoupled from previous star formation history, these objects are either found to lie in regions consistent with the measured main sequence extrapolated to higher \Mstar\ (4/5), or  have very poorly constrained SFR (1/5). These results suggest that the DE SFH did not encompass the true SFH of the objects, leading to the objects being incorrectly interpreted as outliers from the main sequence. We show the postage stamp cut-outs for these objects in Fig.~\ref{fig:panels.png}, which show no strong evidence for the very young ages derived with the DE SFH. 

From so few objects with firm constraints with the more flexible SFH, we cannot comment on the likely bias on the measurements of the main sequence due to our original choice. However, we have demonstrated that fitting with a more complex SFH would have been unfeasible for our sample given the current broad-band constraints without firm priors. \cite{leja_older_2019} used more flexible SFHs in their analysis, but demonstrated that the results are dependent on the priors on their SFH in \cite{leja_how_2019}. They chose a ``continuity'' prior that down-weights sharp transitions between bins of star formation. This prior is somewhat justified by the demonstration that the chosen SFH+priors brings the observed cosmic SFR density and stellar mass growth into agreement for the first time. This means that on average the prior is appropriate at the redshift studied ($z\lesssim2.5$). However, it is not clear that these priors are suitable for determining the impact of bursty star formation on the form of the main sequence at higher redshifts ($z\gtrsim2.5$), where short timescale star formation can significantly affect the broad-band fluxes via the emission lines. Ideally we should first use data to determine how bursty the star formation is in systems at high redshift to settle on a suitable prior. Our demonstration here shows that the data-set used in this study is not appropriate for this purpose for most galaxies in the sample. We must await \JWST\ data-sets, with medium-band filters spanning the rest-frame continuum, and rest-frame optical spectroscopy constraining the emission lines, and even continuum emission with the lower resolution mode. 

\subsection{The limitations of the filter-set and prospects for main sequence measurements in the future}
\label{ss:Poor Constraints at higher redshifts}

The filter-set used in this work is close to the optimum available data to study the low-mass end of the main sequence at high redshifts before the advent of \JWST{}. The addition of HAWK-I Ks-band provides a vital data-point between the reddest \HST{} filter and bluest \Spitzer{} filter. 

Fig.~\ref{fig:emission_lines} displays the filter coverage over key spectral features as a function of redshift for the \astrodeep\ (top panel) and the \JWST\ Advanced Deep Extra-galactic Survey (JADES) filter-set (bottom panel). The Lyman and Balmer breaks are shown as dashed lines, while key emission lines are shown as solid lines with corresponding labels. Regions where the Ks-band or IRAC 3.6 and 4.5 \um{} filters are contaminated by bright emission lines (either \oii{}, \oiii{}, \Hb{} or \Ha{}) are shown as hatched regions. To derive firm \Mstar{} and \sfr{} estimates requires a firm photometric redshift estimate as well as filters sampling the rest-frame ultra-violet to rest-frame optical. In particular, the rest-frame optical should provide constraints of the stellar continuum free from contamination by bright emission lines. One would ideally also have reasonable constraints on the shape of the Balmer break. The firm photometric redshift can come from filters bracketing either the Lyman or Balmer breaks. 

For the \astrodeep\ filter-set, the $3<z<4$ bin has two \Spitzer{} filters probing the rest-frame optical free from emission-line contamination, but the Lyman break is passing through the lowest wavelength filter (F435W) and the Balmer break strength and position is muddied by the contamination of \oiii{} and \Hb{} to the Ks-band. Both these effects significantly reduce the accuracy of the photometric redshift constraints, which impacts the constraints on \Mstar{} and \sfr{}, and also the constraints on \scatter{}, as explained in Section~\ref{sss:redshift_bin_results}. By $z>4$, the Lyman break is securely bracketed by two filters, thereby improving the photometric redshift constraints, and the 4.5 \um{} filter provides constraints of the stellar continuum in the rest-frame optical. This provides firmer \Mstar{} and \sfr{} constraints, and explains why our scatter estimate at $4<z<5$ is no-longer significantly under-estimated in the redshift bin analysis (see Fig.~\ref{fig:redshift_bins}).

The bottom panel of Fig.~\ref{fig:emission_lines} shows the coverage from an example \JWST\ NIRCam filter-set, that was chosen for JADES. The coverage from the two longest wavelength broad-band filters (F356W and F444W) appears similar to that of IRAC 3.5 and 4.5 \um. However, imaging with the F356W and F444W filters will have a far greater depth and resolution, which in turn minimizes the uncertainties arising from the deconfusion process. The F200W and F277W filters provide wavelength coverage across the gap between the \HST\ and \Spitzer\ filters in the \astrodeep\ catalogue. Additionally there are two medium-band filters (F335M and F410M), which in total provides six filters red-ward of the Balmer break, significantly mitigating the issue of emission line contamination. There are considerably more medium and narrow band filters to choose from, and part of the area covered by JADES (specifically the Hubble Ultra Deep Field) will also be visited by proposal 1963 \citep{williams_udf_2021}.  This imaging will provide additional medium band filters (F182M, F210M, F430M, F460M and F480M), further sampling the rest-frame optical of high redshift galaxies.  This extra imaging will provide multiple anchors probing the stellar continuum free from emission line contamination.

It is worth noting that the addition of optical \HST\ ACS photometry is still beneficial in determining robust photometric redshifts where the JADES filter-set does not bracket the Lyman break, as well as for constraining the rest-frame UV continuum at low redshifts. 

\section{Conclusions}
\label{s:Conclusions}

We used \beagle\ to fit to photometry in the \astrodeep{} catalogue for the first four Frontier Field clusters: Abell 2744, MACS0416, MACS0717 and MACS1149. Gravitational lensing due to the large foreground clusters has enabled us to probe masses as low as $10^7-10^8\,\Msun$,  between redshifts $1.25<z<6.0$. 

We have presented a Bayesian hierarchical model of the star-forming main sequence which accounts for the heteroskedastic, co-varying errors on stellar mass, SFR and redshift, as well as the presence of outliers.

To determine a suitable parametrization for our full model, we initially fitted the main sequence relation within different redshift bins. Our initial analysis demonstrated that the sampling of galaxy SEDs provided by the filter-set used (representing the best filter-set currently available for probing faint galaxies at high redshift) provides \Mstar\ and \sfr\ estimates that are too poorly constrained to warrant fitting with a fully flexible model. We describe here the decisions made and results for the redshift-dependent model. 

\begin{itemize}
    \setlength\itemsep{1em}
    \item We fit with a slope that is constant with redshift, measuring $\slope=0.79\pm^{0.03}_{0.04}$.
    
    \item We choose a physically motivated parametrization for the evolution in the normalization of the main sequence, based on the expected evolution of accretion rate of gas onto the parent halos, with the form $\intercept(z)=\log(\ssfrNorm(1+z)^{\ssfrPower})+0.7$.  We measure $\ssfrNorm=0.12^{+0.04}_{-0.03}$ and $\ssfrPower=2.40^{+0.18}_{-0.18}$.  The value of \ssfrPower\ is consistent with the value expected if sSFR scales with accretion onto dark matter halos (a value of 2.25) and the data is consistent with a rising sSFR to high redshifts.
    
    \item Having removed the majority of outliers located below the main sequence due to their highly unconstrained measurements of SFR, we account for outliers at $z<4$ by modelling them simply as belonging to a broad Gaussian distribution in \sfr\ with mean and standard deviation constant with redshift, as well as the probability of an object being an outlier. 
    
    \item For $z>4$ we set the probability of outliers to zero finding no strong evidence for them from the redshift bin results.
    
    \item We find that intrinsic scatter about the main sequence is highly degenerate with the outlier model parameters, and cannot be accurately determined separately within the $3<z<4$ bin. For the full model we resort to fitting a scatter that is constant with redshift, and measure an intrinsic scatter (deconvolved from uncertainties on \Mstar\ and \sfr) of $\scatter=0.26^{+0.02}_{-0.02}$.
    
\end{itemize}

We have explored the limitations of the data and demonstrated how to diagnose when the data may be insufficient to constrain the star-forming main sequence without significant biases. We re-fitted the galaxies in the $1.25<z<2$ bin (those galaxies in our sample with the most complete sampling of their SEDs, and therefore likely the best physical parameter constraints) in two ways.  First, with a different prior on $\tausfr$, which describes the timescale of decay in our delayed exponentially declining SFH.  Our results show that with our fiducial prior, the \Mstar\ and \sfr\ estimates appeared well-constrained, yet when the prior is changed, it shows that objects which were originally fitted with $\log(\sfrInLog/\Msun\,\yr^{-1})\lesssim0.0$ give much lower SFR estimates.  The fiducial prior was therefore somewhat informative and veiling which objects had poorly constrained SFR.  We also re-fit galaxies in the $1.25<z<2$ bin with a less constraining SFH that allowed the recent $10\,\Myr$ of constant star formation to vary independently of the previous SFH.  We demonstrate how few objects had well-constrained \Mstar\ and \sfr\ estimates with this history, meaning that in order to fit more complex and realistic SFHs, we first require an improved data-set with better constraints.

The improved sampling of the SED that can be achieved with \JWST\ NIRCam broad and medium-band filters, as well as the consistent depth that can be achieved will significantly improve the constraints on the main sequence at high redshifts.

\section*{Acknowledgments}

LS and RM acknowledge support by the Science and Technology Facilities Council (STFC) and European Research Council (ERC) Advanced Grant 695671 ``QUENCH''. RM also acknowledges funding from a research professorship from the Royal Society. ECL acknowledges support of
an STFC Webb Fellowship (ST/W001438/1). JC acknowledges funding from the ERC Advanced Grant 789056 ``FirstGalaxies'' (under the European Union’s Horizon 2020 research and innovation programme).

\nocite{smit_evidence_2014, stark_keck_2013, marmol-queralto_evolution_2016, thorne_deep_2021}

\section*{Data availability}

The data and code underlying this article are available at https:\slash\slash github.com\slash ls861\slash M-SFR-Sandles2022. The full \astrodeep{} catalogue is publicly available at http:\slash\slash astrodeep.u-strasbg.fr\slash ff\slash .

\bibliographystyle{mnras}

\bibliography{Paper1_bibtex_009}

\label{lastpage}

\end{document}